\documentclass[prx,amsmath,amssymb,reprint]{revtex4-2}

\usepackage{graphicx}
\usepackage{color}
\usepackage{bm}
\usepackage[hidelinks]{hyperref}
\usepackage{siunitx}

\renewcommand*{\vec}[1]{\mathbf{#1}}
\newcommand*{\diff}[1]{\mathrm{d}#1\,}
\newcommand*{\Fexc}{F_\mathrm{exc}}
\newcommand*{\Vext}{V_\mathrm{ext}}

\graphicspath{{./figures/}}

\begin{document}

\title{Neural functional theory for inhomogeneous fluids: Fundamentals and applications}

\author{Florian Sammüller}
\author{Sophie Hermann}
\author{Daniel de las Heras}
\author{Matthias Schmidt}
\email{Matthias.Schmidt@uni-bayreuth.de}
\affiliation{Theoretische Physik II, Physikalisches Institut, Universität Bayreuth, D-95447 Bayreuth, Germany}

\begin{abstract}
  We present a hybrid scheme based on classical density functional theory and machine learning for determining the equilibrium structure and thermodynamics of inhomogeneous fluids.
  The exact functional map from the density profile to the one-body direct correlation function is represented locally by a deep neural network.
  We substantiate the general framework for the hard sphere fluid and use grand canonical Monte Carlo simulation data of systems in randomized external environments during training and as reference.
  Functional calculus is implemented on the basis of the neural network to access higher-order correlation functions via automatic differentiation and the free energy via functional line integration.
  Thermal Noether sum rules are validated explicitly.
  We demonstrate the use of the neural functional in the self-consistent calculation of density profiles.
  The results outperform those from state-of-the-art fundamental measure density functional theory.
  The low cost of solving an associated Euler-Lagrange equation allows to bridge the gap from the system size of the original training data to macroscopic predictions upon maintaining near-simulation microscopic precision.
  These results establish the machine learning of functionals as an effective tool in the multiscale description of soft matter.
\end{abstract}

\maketitle

\section{Introduction}

The problem with density functional theory (DFT) is that you do not know the density functional.
Although this quip by the late and great Yasha Rosenfeld \cite{Evans2002} was certainly meant in jest to a certain degree, it does epitomize a structural assessment of classical DFT \cite{Evans1979,Evans1992,Hansen2013,Evans2016}.
As a general formulation of many-body statistical physics, the framework comprises a beautiful and far reaching skeleton of mathematical formalism centered around a formally exact variational minimization principle \cite{Evans1979,Mermin1965}.
In practice however, the theory needs to be fleshed out by approximations of all means conceivable in our efforts to get to grips with the coupled many-body problem that is under consideration.
Specifically, it is the excess (over ideal gas) intrinsic Helmholtz free energy $\Fexc[\rho]$, expressed as a functional of the position-resolved density profile $\rho(\vec{r})$, which needs to be approximated.

Decades of significant theoretical efforts have provided us with a single exact functional, that for nonoverlapping hard rods in one spatial dimension, as obtained by another hero in the field, Jerry Percus \cite{Percus1976}.
Nevertheless, useful DFT approximations range from the local density approximation for large scale features which are decoupled from microscopic length scales, to square-gradient functionals with their roots in the 19th century, to the arguably most important modern development, that of the fundamental measure theory (FMT) as kicked off by Rosenfeld in 1989 \cite{Rosenfeld1989} and much refined ever since \cite{Kierlik1990,Kierlik1991,Phan1993,Roth2002,HansenGoos2006,Roth2010,Tarazona2000,Tarazona2008}.
FMT is a geometry-based framework for the description of hard sphere systems and it has deep roots in the Percus-Yevick \cite{Percus1958} and scaled-particle theories \cite{Hansen2013}, which Rosenfeld was able to unify and generalize based on his unique theoretical insights \cite{Rosenfeld1988}.

The realm of soft matter \cite{Nagel2017,Evans2019,Barrat2003} stretches far beyond the hard sphere fluid.
FMT remains relevant though in the description of a reference system as used e.g.\ in studies of hydrophobicity, where the behaviour of realistic water models \cite{Molinero2008,Coe2022} is traced back to the simpler Lennard-Jones fluid, which in turn is approximated via the hard sphere FMT functional plus a mean-field contribution for describing interparticle attraction \cite{Evans2019,Coe2022a,Coe2023}.
Further topical uses of FMT include the analysis of the three-dimensional electrolyte structure near a solid surface \cite{MartinJimenez2016,HernandezMunoz2019} and the problem of the decay length of correlations in electrolytes \cite{Cats2021}.

There is a current surge in the use of machine learning techniques in soft matter, e.g.\ for its characterization \cite{Clegg2021}, engineering of self-assembly \cite{Dijkstra2021}, structure detection \cite{Boattini2019}, and for learning many-body potentials \cite{CamposVillalobos2021,CamposVillalobos2022}.
Within classical DFT, machine learning was used to address ordering of confined liquid crystals \cite{SantosSilva2014}, and free energy functionals were obtained for one-dimensional systems from convolutional \cite{Lin2019} and equation-learning \cite{Lin2020} networks as well as within a Bayesian inference approach \cite{Yatsyshin2022}.
Cats \textit{et~al.}~\cite{Cats2021a} used machine learning to improve the standard mean-field approximation of the excess Helmholtz free-energy functional for the Lennard-Jones fluid.
In nonequilibrium, de las Heras \textit{et~al.}~\cite{Heras2023} have reported a method to machine-learn the functional relationship of the local internal force for a steady uniaxial compressional flow of a Lennard-Jones fluid at constant temperature.
As prescribed by power functional theory \cite{Schmidt2013,Schmidt2022}, the functional dependence in nonequilibrium not only incorporates the density profile but also the one-body current.

In this work, we return to the problem of describing and predicting the structure and thermodynamics of inhomogeneous equilibrium fluids.
We show that a neural network can be trained to accurately represent the functional dependence of the one-body direct correlation function with respect to the density profile.
The presented methods are directly applicable to virtually arbitrary fluids with short-ranged interparticle interactions.
In the following, we focus on the well-studied hard sphere fluid in order to exemplify our framework and to challenge the available highly accurate analytic approaches from liquid integral equation theory and FMT.
We give more details about the feasibility of generalizations in the discussion.
Reference data for training and testing the model is provided by grand canonical Monte Carlo (GCMC) simulations that cover a broad range of randomized inhomogeneous environments in planar geometry.

We implement functional calculus on the basis of the trained neural functional to infer related physical quantities and demonstrate their consistency with known literature results both in bulk and in inhomogeneous systems.
In particular, we highlight the accessibility of the fluid pair structure, the determination of free energies and equations of state as well as the validation of thermal Noether sum rules \cite{Hermann2021}.
These results corroborate that the neural functional exceeds its role as a mere interpolation device and instead possesses significant representational power as a genuine density functional for the prediction of nontrivially related physical properties.
We apply the trained neural network in the DFT Euler-Lagrange equation, which enables the self-consistent calculation of density profiles and which hence constitutes a neural-network-based DFT or short neural DFT.
This method alleviates conventional DFT from the burden of having to find suitable analytic approximations while still surpassing even the most profound existing treatments of the considered hard sphere fluid via FMT functionals \cite{Rosenfeld1989,HansenGoos2006,Roth2010} in accuracy.
We further demonstrate the fitness of the method for the straightforward application to multiscale problems.
Neural DFT therefore provides a way to transfer near-simulation microscopic precision to macroscopic length scales, which serves as a technique to predict properties of inhomogeneous systems which far exceed typical box sizes of the original training data.

This work is structured as follows.
The relevant physical background of liquid state theory is provided in Sec.~\ref{sec:physics}.
Details of the simulations as well as of the neural network are given in Secs.~\ref{sec:simulation} and \ref{sec:neural_network}.
The training procedure and results for the achieved metrics that measure its convergence are presented in Sec.~\ref{sec:training}.
We proceed by testing physical properties of the trained model and use automatic differentiation of the neural network in Sec.~\ref{sec:twobody_bulk} to access pair correlations, which are then compared to bulk results from both the Percus-Yevick theory and from simulations.
The consistency of the neural direct correlation functional to satisfy thermal Noether sum rules is validated in Sec.~\ref{sec:noether}, and different ways to obtain the bulk equation of state as well as free energies in inhomogeneous systems are given in Sec.~\ref{sec:free_energy}.
In Sec.~\ref{sec:beyond_analytic}, we show the application of the neural functional to the self-consistent calculation of density profiles via the DFT Euler-Lagrange equation and describe the technical details and conceptual advantages of this neural DFT over analytic approaches.
In Sec.~\ref{sec:beyond_fmt}, the results are compared to those from FMT, and in Sec.~\ref{sec:beyond_the_box}, the relevance of the method for making macroscopic predictions is illustrated for cases of randomized external potential and for sedimentation between hard walls on length scales that far exceed the training simulation box sizes.
We conclude with a discussion of the results and give an outlook to possible improvements and extensions of the method as well as to its application for different fluid types, in more general geometries and in nonequilibrium.

\section{Results}

\subsection{Machine learning intrinsic correlations}

\subsubsection{Physical background}
\label{sec:physics}

We start with the standard relation for the one-body direct correlation function $c_1(\vec{r})$ of liquid state theory \cite{Hansen2013},
\begin{equation}
  \label{eq:c1_EL}
  c_1(\vec{r}) = \ln \rho(\vec{r}) + \beta \Vext(\vec{r}) - \beta \mu,
\end{equation}
where $\vec{r}$ denotes the spatial position and $\beta = 1 / (k_B T)$ with the Boltzmann constant $k_B$ and absolute temperature $T$.
The three terms on the right hand side of Eq.~\eqref{eq:c1_EL} represent respectively the ideal gas contribution, the external potential $\Vext(\vec{r})$ and the influence of the particle bath at chemical potential $\mu$.
The logarithm in Eq.~\eqref{eq:c1_EL} is understood as $\ln[\Lambda^3 \rho(\vec{r})]$ with the thermal wavelength $\Lambda$, which can be set to the particle size $\sigma$ without any loss of information in the present classical context.
For a prescribed external potential $\Vext(\vec{r})$, knowledge of the corresponding equilibrium density profile $\rho(\vec{r})$ allows to compute $c_1(\vec{r})$ explicitly via Eq.~\eqref{eq:c1_EL}.
This relationship can be viewed as a locally resolved chemical potential balance: the contribution from the ideal gas, $k_B T \ln \rho(\vec{r})$, from the external potential, $\Vext(\vec{r})$, and from interparticle interactions, $- k_B T c_1(\vec{r})$, add up at each position to $\mu$, which is necessarily uniform throughout an equilibrium system.

However, the notation in Eq.~\eqref{eq:c1_EL} is oblivious to a central result shown by Evans \cite{Evans1979} in 1979, thereby kicking off a modern theory for the description of inhomogeneous fluids.
For given type of internal interactions, the spatial variation of the function $c_1(\vec{r})$ is already uniquely determined by the spatial form of the density profile $\rho(\vec{r})$ alone, without the need to invoke the external potential explicitly.
From this vantage point of classical DFT, the dependence of $c_1(\vec{r})$ on $\rho(\vec{r})$ is not merely pointwise but rather with respect to the values of the entire density profile, which determine $c_1(\vec{r})$ at each given position $\vec{r}$.
Formally, this relationship is exact \cite{Hansen2013,Evans1979} and it constitutes a functional dependence $c_1(\vec{r}; [\rho])$, which is indicated by brackets here and in the following and which is in general nonlinear and nonlocal.
As we will demonstrate, the existence of such a universal functional mapping makes the problem of investigating inhomogeneous fluids particularly amenable to supervised machine learning techniques.

In most formulations of classical DFT, one exploits the fact that the intrinsic excess free energy functional $\Fexc[\rho]$ acts as a functional generator such that the one-body direct correlation function is obtained via functional differentiation with respect to the density profile,
\begin{equation}
  \label{eq:c1_Fexc}
  c_1(\vec{r}; [\rho]) = - \frac{\delta \beta \Fexc[\rho]}{\delta \rho(\vec{r})}.
\end{equation}
A compact description of standard formulae for the calculation of functional derivatives can be found in Ref.~\cite{Schmidt2022}.
In order to make progress in concrete applications, one typically needs to rely on using an approximate form of $\Fexc[\rho]$ for the specific model under consideration, as determined by its interparticle interactions.
DFT is a powerful framework, as using $c_1(\vec{r}; [\rho])$ obtained from Eq.~\eqref{eq:c1_Fexc} with a suitable expression for $\Fexc[\rho]$ turns Eq.~\eqref{eq:c1_EL} into an implicit equation for the equilibrium density profile $\rho(\vec{r})$.
In the presence of a known form of $\Vext(\vec{r})$, one can typically solve Eq.~\eqref{eq:c1_EL} very efficiently, allowing ease of parameter sweeps, e.g.\ for exhaustive phase diagram explorations.
On the downside, $\Fexc[\rho]$ and thus also $c_1(\vec{r}; [\rho])$ remain approximate and the development of analytic tools has certainly slowed down over several years if not decades.

Here we proceed differently and bypass the excess free energy functional $\Fexc[\rho]$ at first.
Instead, we use a deep neural network to learn and to represent the functional relationship $\rho(\vec{r}) \rightarrow c_1(\vec{r})$ directly, which has significant advantages both for the generation of suitable training data as well as for the applicability of the model in the determination of fluid equilibria.
This investigation is based on GCMC simulations that serve to provide training, validation and test data.
Discriminating between these three roles of use is standard practice in machine learning and we give further details below.

\subsubsection{Simulation method}
\label{sec:simulation}

Generating the simulation data is straightforward and we use the following strategy, adopted to planar situations where the position-dependence is on a single position variable $x$ while the system remains translationally invariant in the $y$- and $z$-direction.
This geometry is highly relevant to identify the physics in planar capillary and adsorption situations and facilitates ease of accurate sampling.
We employ randomized simulation conditions by generating external potentials of the form
\begin{equation}
  \label{eq:Vext}
  \Vext(x) = \sum_{n=1}^4 A_n \sin\left(\frac{2 \pi n x}{L} + \phi_n\right) + \sum_n V_n^\mathrm{lin}(x),
\end{equation}
where $A_n$ and $\phi_n$ are randomly selected Fourier coefficients and phases, respectively, and $L$ is the simulation box length in $x$-direction.
The phases $\phi_n$ are chosen uniformly in the interval $[0, 2\pi)$ and values of $A_n$ are drawn from a normal distribution with zero mean and variance $2.5$.
We choose $L = 20 \sigma$, although there is no specific compliance requirement for the neural network (see below), and the lateral box lengths are set to $10\sigma$ to minimize finite-size effects.
Periodic boundary conditions apply in all spatial directions.
The sinusoidal terms in $\Vext(x)$ are complemented by up to five piecewise linear functions $V^\mathrm{lin}(x) = V_1 + (V_2 - V_1) (x - x_1) / (x_2 - x_1)$ for $x_1 < x < x_2$ and 0 otherwise, for which the parameters $0 < x_1 < x_2 < L$, $V_1$ and $V_2$ are again chosen randomly.
The locations $x_1$ and $x_2$ are distributed uniformly while $V_1$ and $V_2$ follow again from an unbiased normal distribution with variance $4$.
Additionally to the discontinuous linear segments, we explicitly impose planar hard walls in a subset of the simulations by setting $\Vext(x) = \infty$ for $x < x_w/2$ and $x > L - x_w/2$, i.e.\ near the borders of the simulation domain; the width $x_w$ of the wall is chosen randomly in the interval $1 \leq x_w / \sigma \leq 3$.
To cover a broad range from dilute to dense systems, the chemical potential is chosen uniformly within the range $-5 \leq \beta \mu \leq 10$ for each respective GCMC simulation run.
The observed mean densities range from $0.006 \sigma^{-3}$ to $0.803 \sigma^{-3}$, yet smaller and much larger local densities occur due to the inhomogeneous nature of the systems.

In total, 750 such GCMC runs are used, where for given form of $\Vext(x)$ the planar one-body profiles $\rho(x)$ and $c_1(x)$ are obtained.
The former is acquired from straightforward histogram filling and the latter from evaluating Eq.~\eqref{eq:c1_EL} on the basis of the sampled histogram for $\rho(x)$ as well as the known form of $\Vext(x)$ and value of $\mu$ for the specific run under consideration.
As Eq.~\eqref{eq:c1_EL} is undefined for vanishing density, we have excluded regions where $\rho(x) = 0$ such as within the hard walls.
By modern standards of computational resources, the workload for the generation of the simulation data is only moderate at a total CPU time of $\sim 10^4$ hours.

\subsubsection{Neural network}
\label{sec:neural_network}

\begin{figure}[tb]
  \centering
  \includegraphics{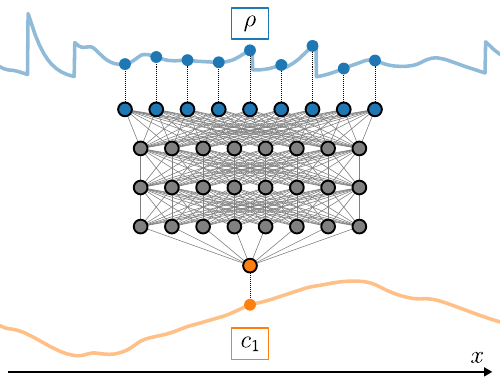}
  \caption{
    We represent the functional mapping from the density profile $\rho(x)$ to local values of the one-body direct correlation function $c_1(x)$ in planar geometry via a neural network.
    The density profile is discretized on a regular spatial grid with resolution $0.01 \sigma$ and given within a region around the location of interest to the input layer.
    Three fully-connected layers with continuously differentiable activation functions enable the inference of the nonlinear and nonlocal functional map.
    The output layer consists of a single node which yields the predicted value of $c_1(x)$ at the chosen location.
  }
  \label{fig:network}
\end{figure}

We use a deep neural network \cite{Chollet2017} to represent the functional map from the density profile to the \emph{local} value of the one-body direct correlation function at a given point.
That is, instead of the entire function, we construct the network to output only the scalar value $c_1(x)$ for a certain position $x$ when supplied with the surrounding inhomogeneous density.
The relevant section of the density profile comprises the values of $\rho(x)$ in a specified window around a considered location $x$, as described below.
Despite the locality of the method, access to the entire (discretized) one-body direct correlation profile is immediate via evaluation of the neural network at pertinent positions $x$ across the domain of interest.
Multiple local evaluations of the network remain performant on highly parallel hardware such as GPUs when passing the input accordingly in batches.
A schematic picture of the network architecture is given in Fig.~\ref{fig:network} and is explained in the following.

The functional dependence on the density profile is realized by providing discretized values of $\rho(x)$ on an equidistant grid with resolution $\Delta x = 0.01 \sigma$.
As $c_1(x; [\rho])$ depends only on the immediately surrounding density profile around a fixed location $x$, we restrict the input range $x'$ to a sufficiently large window $x' \leq |x - x_c|$.
We choose the cutoff $x_c = 2.56 \sigma$ based on simulation data for the bulk direct correlation function \cite{Groot1987} and on the evaluation of training metrics for different window sizes $x_c$.
Increasing the value of $x_c$ further led to no improvement in the performance of the trained neural network.

This behavior is expected from theoretical considerations, as the one-body direct correlation function vanishes quickly for short-ranged pair potentials \cite{Hansen2013}.
We recall that in FMT, $x_c = \sigma$ by construction.
Note that the choice of $c_1(x; [\rho])$ as our target functional is not coincidental, but that its quick spatial decay rather is a pivotal characteristic central to the success of our method.
To contrast this, assume that one attempts to model the functional mapping $\mu_\mathrm{loc}(x) = \mu - \Vext(x) \rightarrow \rho(x)$, thereby naively imitating the simulation procedure.
This task poses major challenges due to the long-range nature of density correlations induced by an external potential, which is circumvented in our case by the choice of a more manageable target functional.

The input layer involves 513 nodes and is followed by three fully-connected hidden layers with 512 units each.
The output layer consists of a single node for the scalar value of $c_1(x)$ at the specified location $x$.
In order to realize a nonlinear input-output mapping of the neural network, activation functions are applied to the output of each node within a hidden layer (see also Ref.~\cite{Chollet2017} for a pedagogical introduction to the design of neural networks).
We deviate here from the most common choice of a rectified linear unit (ReLU) and instead employ continuously differentiable activation functions such as the exponential linear unit or the softplus function \cite{Dubey2022}.
This choice leads to substantial improvements during training and in particular when using automatic differentiation to evaluate two-body quantities, see Secs.~\ref{sec:twobody_bulk} and \ref{sec:noether}.
We attribute the superior performance to the fact that activation functions which are not continuously differentiable and which vanish in certain domain ranges (such as ReLU) reinforce sparsity of the activation output, i.e.\ the tendency to set many units of a hidden layer identically to zero \cite{Glorot2011}.
While this property is desired in many machine learning tasks (e.g.\ for classification), it hinders the accurate representation of the functional relation $c_1(x; [\rho])$ in our case.
The resulting neural functional for the one-body direct correlation function is denoted in the following by $c_1^\star(x; [\rho])$ and related quantities which follow from it by inference are marked accordingly by a superscript star.

\subsubsection{Training procedure and metrics}
\label{sec:training}

The machine learning routines are implemented in Keras/Tensorflow \cite{Chollet2017} and we use the standard Adam \cite{Kingma2014} optimizer for the adjustment of the network parameters in order to fit $c_1^\star(x; [\rho])$ against the simulation reference $c_1(x)$.
The problem at hand is a regression task.
Hence, the mean squared error is chosen as a suitable loss function and the mean average error serves as a validation metric.
Since the model shall infer the pointwise value $c_1(x)$ from a density section around a specified location $x$, see Fig.~\ref{fig:network}, the simulation data cannot be passed as is to the neural network.
Instead, windowed views of the density profile have to be generated prior to the training loop, which correspond to the target value $c_1(x)$ at the center $x$ of the respective window.
A periodic continuation of all simulation profiles is valid due to periodic boundary conditions.
Additionally, we use data augmentation to benefit from the inherent mirror symmetry (i.e.\ $x \rightarrow -x$) of the problem and thus effectively double the number of training data sets.
As is customary, we separate the independent simulation results prior to performing the machine learning routines: 150 are kept aside as a test set, 150 serve as validation data to monitor training progress and 450 are used for the actual training of the neural network.

Modeling the functional relationship of $c_1(x; [\rho])$ \emph{locally}, i.e.\ inferring pointwise values individually instead of outputting the entire profile at once, has numerous conceptual and practical advantages.
Regarding the feasibility of the neural network in concrete applications, one is free to choose an arbitrary box length $L$ when gathering training data and more importantly to readjust the value of $L$ when using the trained neural network for making predictions (cf.\ Sec.~\ref{sec:beyond_the_box}).
From a physical point of view, providing only local density information has the merit of already capturing the correlated short-range behavior of $c_1(x; [\rho])$.
If the neural network were to output the entire one-body direct correlation profile from a given density profile $\rho(x)$ at once, this inherent locality would have to be learned instead, hence leading to a much more elaborate training process.
Lastly, the fine-grained nature of the training data turns out to be highly beneficial from a machine learning perspective.
Note that one can generate $9 \cdot 10^5$ input-output pairs from 450 training simulations in the present context (with the values being doubled after data augmentation).
The increased cardinality of the training set enables better generalization of the model and also prevents overfitting, e.g.\ to the statistical noise of the sampled profiles.

We train the model for 100 epochs in batches of size 256 and decrease the learning rate exponentially by $\sim 5\%$ per epoch from an initial value of $0.001$.
This results in a best mean average error of $0.0022$ over the validation set, which is of the same order as the estimated average noise of the simulation data for $c_1(x)$.
Therefore, we deem our neural network to possess full representational power of the local functional relationship $c_1(x; [\rho])$ within the conditions of the provided simulation data.
Code, simulation data and trained models are published online \cite{Zenodo}.

\subsection{Examining the neural correlation functional}

\subsubsection{Two-body bulk correlations}
\label{sec:twobody_bulk}

Besides monitoring standard metrics such as the mean average error over a test set, arguably deeper physical insights into the rigorous structure of the statistical mechanics at hand serves for assessing the quality of the neural functional $c_1^\star(x; [\rho])$.
We first ascertain that the model gives an accurate representation of the physics of bulk fluids.
Despite the apparent simplicity of this case, this is a highly nontrivial test as the training data solely covered (strongly) inhomogeneous situations.
For this, we investigate the pair structure and aim at implementing the two-body direct correlation functional, which is formally defined as the functional derivative \cite{Hansen2013}
\begin{equation}
  \label{eq:c2}
  c_2(\vec{r}, \vec{r}'; [\rho]) = \frac{\delta c_1(\vec{r}; [\rho])}{\delta \rho(\vec{r}')}.
\end{equation}
On the basis of the neural network, we can make use of the powerful automatic differentiation techniques.
This allows to create an immediate analog of Eq.~\eqref{eq:c2} via $c_2^\star(x, x'; [\rho]) = \delta c_1^\star(x; [\rho]) / \delta \rho(x')$, where the functional derivative $\delta / \delta \rho(x')$ is evaluated by reverse mode automatic differentiation with respect to the input values of the discretized density profile.
In common machine learning frameworks, this requires only high-level code (e.g.~\verb|GradientTape| in Keras/Tensorflow \cite{Chollet2017}).
The numerical evaluation of $c_2^\star(x, x'; [\rho])$ is performant as reverse mode automatic differentiation generates executable code that is suitable for building derivatives with respect to multiple input variables simultaneously.

\begin{figure}[!tb]
  \centering
  \includegraphics{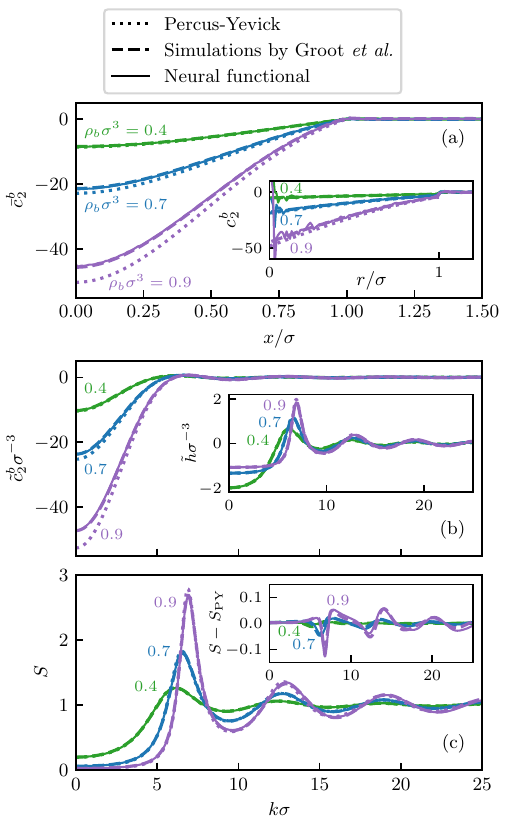}
  \caption{
    We compare (a) the planar direct correlation function $\bar{c}_2^b(x)$, (b) its radial Fourier space representation $\tilde{c}_2^b(k)$, and (c) the static structure factor $S(k)$ for different bulk densities $\rho_b \sigma^3 = 0.4, 0.7, 0.9$ (as indicated).
    Data is shown as obtained from the Percus-Yevick theory (dotted), from simulation results by Groot \textit{et~al.}~\cite{Groot1987} (dashed) and from our neural functional $c_1^\star(x; [\rho])$ (solid), where $\bar{c}_2^{b \star}(x)$ is acquired via automatic differentiation.
    The inset in panel (a) shows the radial direct correlation function $c_2^b(r)$ as obtained via Eq.~\eqref{eq:c2_planar_to_radial}.
    In panel (b), the inset depicts the total correlation function $\tilde{h}(k)$ in Fourier space, which follows from $\tilde{c}_2^b(k)$ via the bulk Ornstein-Zernicke Eq.~\eqref{eq:OZ_Fourier}.
    The inset in panel (c) displays the deviation of $S(k)$ to the Percus-Yevick result $S_\mathrm{PY}(k)$ for the simulation data and the neural functional.
    Simulation and machine learning results are in very good agreement with each other while the Percus-Yevick theory shows quantitative discrepancies.
  }
  \label{fig:twobody_bulk_comparison}
\end{figure}

We obtain the bulk direct correlation function in planar geometry as the special case $\bar{c}_2^b(x, \rho_b) = c_2(0, x; [\rho_b])$, where we have introduced the bulk density $\rho_b(x) = \rho_b = \mathrm{const}$.
(In the notation, the parametric dependence on $\rho_b$ is dropped in the following.)
Note that $\bar{c}_2^b(x)$ is distinct from the more common radial representation $c_2^b(r)$, as our geometry implies an integration over the lateral directions $y$ and $z$, i.e.\
\begin{equation}
  \label{eq:c2_radial_to_planar}
  \begin{split}
    \bar{c}_2^b(x) &= \int \diff{y} \diff{z} c_2^b\left(r = \sqrt{x^2 + y^2 + z^2}\right)\\
                   &= 2 \pi \int_x^\infty \diff{r} r c_2^b(r),
  \end{split}
\end{equation}
where the last equality follows from using radial coordinates and substitution.
The standard radial form $c_2^b(r)$ can however be recovered by differentiating Eq.~\eqref{eq:c2_radial_to_planar} with respect to $x$ such that
\begin{equation}
  \label{eq:c2_planar_to_radial}
  c_2^b(r) = -\frac{\bar{c}_2^{b \prime}(r)}{2 \pi r},
\end{equation}
where $\bar{c}_2^{b \prime}(r)$ denotes the derivative of $\bar{c}_2^b(x)$ evaluated at $x = r$.
Numerical artifacts might occur particularly for small values of $r$ as evaluating Eq.~\eqref{eq:c2_planar_to_radial} requires the numerical derivative of $\bar{c}_2^b(x)$ as well as a division by $r$.

We perform a Fourier transform of the planar real space representation $\bar{c}_2^b(x)$ and utilize radial symmetry in Fourier space.
This acts as a deconvolution of Eq.~\eqref{eq:c2_radial_to_planar} and directly yields the radial Fourier (Hankel) transform of $c_2^b(r)$,
\begin{equation}
  \label{eq:c2_bulk_Fourier}
  \tilde{c}_2^b(k) = \frac{4 \pi}{k} \int_0^\infty \diff{r} r \sin(kr) c_2^b(r).
\end{equation}
The inverse transform is identical to Eq.~\eqref{eq:c2_bulk_Fourier} up to a factor of $(2 \pi)^{-3}$ upon interchanging $r$ and $k$.
To go further, the bulk Ornstein-Zernike equation \cite{Hansen2013}
\begin{equation}
  \label{eq:OZ_Fourier}
  \tilde{c}_2^b(k) = \frac{\tilde{h}(k)}{1 + \rho_b \tilde{h}(k)}
\end{equation}
is used to obtain the total correlation function $\tilde{h}(k)$ from $\tilde{c}_2^b(k)$ in Fourier space after rearrangement.
Recall that the radial distribution function follows directly via $g(r) = h(r) + 1$; here $h(r)$ is the real space representation of $\tilde{h}(k)$.
The static structure factor $S(k)$ is then given as
\begin{equation}
  \label{eq:structure_factor}
  S(k) = 1 + \rho_b \tilde{h}(k).
\end{equation}

In Fig.~\ref{fig:twobody_bulk_comparison}, results of $\bar{c}_2^b(x)$, $c_2^b(r)$, $\tilde{c}_2^b(k)$, $\tilde{h}(k)$ and $S(k)$ are shown for different bulk densities $\rho_b \sigma^3 = 0.4, 0.7, 0.9$.
From our neural functional, we obtain $\bar{c}_2^{b \star}(x) = \delta c_1^\star(0; [\rho]) / \delta \rho(x) \vert_{\rho = \rho_b}$, i.e.\ the autodifferentiated network is evaluated at spatially constant density $\rho_b$.
The total correlation function and the static structure factor follow from Eqs.~\eqref{eq:OZ_Fourier} and \eqref{eq:structure_factor} after having computed $\tilde{c}_2^{b \star}(k)$ via a numerical Fourier transform of $\bar{c}_2^{b \star}(x)$.
For comparison, we also depict reference data obtained analytically from the Percus-Yevick theory \cite{Percus1958} and reproduced from simulation results of Groot \textit{et~al.}~\cite{Groot1987}.
Good agreement is found between simulation and the autodifferentiated neural network, while the Percus-Yevick result shows noticeable deviations in $\bar{c}_2^b(x)$.
The latter overestimates the depth of the core region $x < \sigma$ and this discrepancy increases for larger bulk densities.
The neural functional yields a clear improvement over the Percus-Yevick theory and shows only marginal differences to the simulation results of Ref.~\cite{Groot1987} for both the planar real space and the radial Fourier space representation of the two-body direct correlation function.
In $\tilde{h}(k)$ and $S(k)$, the severity of the discrepancies of simulation and machine learning data to the Percus-Yevick results decreases, but a difference is still noticeable in particular for large bulk densities.
A slight mismatch to the simulation reference is observed in the magnitude and phase of the oscillations of the Percus-Yevick static structure factor $S_\mathrm{PY}(k)$, and this correction is reproduced very well by the neural functional.
Note that although one arrives at radial representations of the quantities $\tilde{h}(k)$ and $S(k)$ in Fourier space, performing the radial backtransform to real space numerically according to the inverse of Eq.~\eqref{eq:c2_bulk_Fourier} is generally a ``notoriously difficult task'' \cite{Henderson1975} and is not considered here.

This successful test reveals that, while being trained solely with one-body profiles, the neural functional $c_1^\star(x; [\rho])$ contains full two-body information equivalent in bulk to the radial distribution function $g(r)$.
The pair correlations can be accessed via automatic differentiation at low computational cost and they are consistent with known bulk results.
We recall that this is a mere byproduct of the neural network and that no such two-body information has been explicitly incorporated in the training.
More so, Fig.~\ref{fig:twobody_bulk_comparison} demonstrates that the bulk quantities $\bar{c}_2^b(x)$, $\tilde{c}_2^b(k)$, $\tilde{h}(k)$ and $S(k)$ as obtained from $c_1^\star(x; [\rho])$ substantially outperform the Percus-Yevick theory and almost attain simulation quality.
In Appendix \ref{appendix:c3}, we illustrate that higher-order correlations such as the three-body direct correlation functional $c_3^\star(x, x', x''; [\rho])$ follow analogously via nested automatic differentiation.
On this level, differences to FMT results are even more prominent than the deviations to the two-body Percus-Yevick results.
As we will show in Sec.~\ref{sec:beyond_fmt}, the accuracy of predictions from the neural network also holds in inhomogeneous situations, where FMT serves again as an analogous and arguably even more challenging theoretical baseline than the Percus-Yevick bulk theory.
Before doing so, we lay out additional consistency tests and quality assessments that are applicable in inhomogeneous systems.

\subsubsection{Noether sum rules}
\label{sec:noether}

\begin{figure}[!tb]
  \centering
  \includegraphics{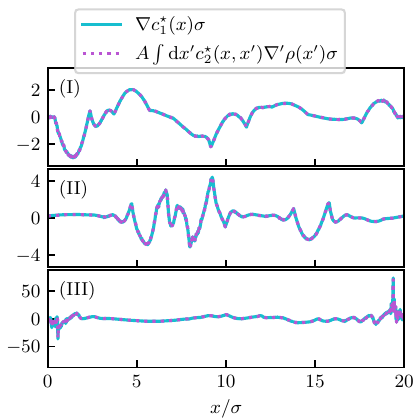}
  \includegraphics{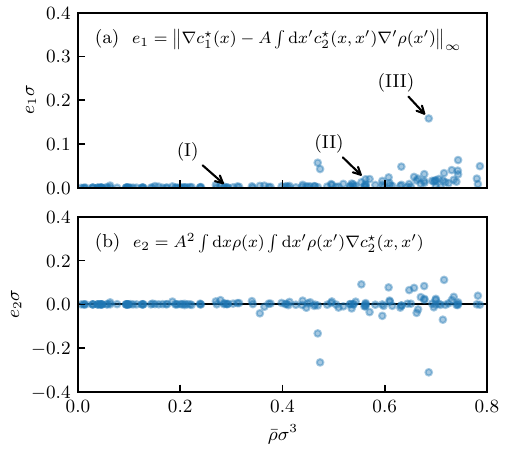}
  \caption{
    Typical profiles of the right and left hand sides of Eq.~\eqref{eq:Noether_correlation} are shown for three test scenarios in panels (I), (II), and (III), where one can verify their high level of agreement across the entire inhomogeneous systems.
    Additionally, the respective scalar discrepancies $e_1$ and $e_2$ of the Noether identities \eqref{eq:Noether_correlation} and \eqref{eq:Noether_shift} are displayed, defined as (a) the maximum norm of the difference of left and right hand side of Eq.~\eqref{eq:Noether_correlation} and (b) the value of the left hand side of Eq.~\eqref{eq:Noether_shift}.
    Across all mean densities $\bar{\rho}$ of the test set, the sum rules are satisfied to very high accuracy by our model.
    Some outliers remain which we attribute to the numerical computation of spatial gradients appearing in Eqs.~\eqref{eq:Noether_correlation} and \eqref{eq:Noether_shift}, see also panel (III) for an example of the noise that this introduces in the respective terms of Eq.~\eqref{eq:Noether_correlation} particularly in the vicinity of hard walls.
  }
  \label{fig:noether}
\end{figure}

In order to further elucidate whether $c_1^\star(x; [\rho])$ quantitatively reproduces fundamental properties of equilibrium many-body systems, we make use of exact sum rules that follow from thermal Noether invariance \cite{Hermann2021}:
\begin{equation}
  \label{eq:Noether_correlation}
  \nabla c_1(\vec{r}) = \int \diff{\vec{r}'} c_2(\vec{r}, \vec{r}') \nabla' \rho(\vec{r}'),
\end{equation}
\begin{equation}
  \label{eq:Noether_shift}
  \int \diff{\vec{r}} \rho(\vec{r}) \int \diff{\vec{r}'} \rho(\vec{r}') \nabla c_2(\vec{r}, \vec{r}') = 0.
\end{equation}
Both Eq.~\eqref{eq:Noether_correlation} and Eq.~\eqref{eq:Noether_shift} apply in any equilibrated inhomogeneous system regardless of the type of internal interactions.
While the interparticle interaction potential does not appear explicitly in Eqs.~\eqref{eq:Noether_correlation} and \eqref{eq:Noether_shift}, it nevertheless determines the functionals $c_1(\vec{r}; [\rho])$ and $c_2(\vec{r}, \vec{r}'; [\rho])$.
Recall that the spatial gradient of the one-body direct correlation function can be identified with the internal equilibrium force profile, $\vec{f}_\mathrm{int}(\vec{r}) = k_B T \nabla c_1(\vec{r})$ \cite{Schmidt2022}.

We verify that the neural functional complies with the above sum rules \eqref{eq:Noether_correlation} and \eqref{eq:Noether_shift} as follows.
Analogous to Sec.~\ref{sec:twobody_bulk}, we use autodifferentiation to evaluate Eq.~\eqref{eq:c2}, but this time retain the full inhomogeneous structure of $c_2^\star(x, x'; [\rho])$.
The left hand side of Eq.~\eqref{eq:Noether_correlation} is obtained straightforwardly from simple evaluation of the neural functional and numerical spatial differentiation.
As input for $\rho(x)$, we use the simulated density profiles of the test set.
Care is required when evaluating the spatial gradients $\nabla \rho(x)$, $\nabla c_1^\star(x; [\rho])$ and $\nabla c_2^\star(x, x'; [\rho])$ due to the amplification of undesired noise, which we reduce by applying a low-pass filter after having taken the numerical derivatives.
The volume integrals reduce in planar geometry to $\int \diff{\vec{r}} = A \int \diff{x}$, where $A$ is the lateral system area.

In Fig.~\ref{fig:noether}, three typical profiles for the left and right hand side of Eq.~\eqref{eq:Noether_correlation} are shown.
In all three systems both sides of the equation coincide up to numerical noise due to the required spatial derivatives.
Additionally, we define errors via scalar deviations from equality in Eqs.~\eqref{eq:Noether_correlation} and \eqref{eq:Noether_shift} respectively as
\begin{align}
  \label{eq:Noether_errors}
  e_1 &= \left\Vert \nabla c_1(x) - A \int \diff{x'} c_2(x, x') \nabla' \rho(x') \right\Vert_\infty,\\
  e_2 &= A^2 \int \diff{x} \rho(x) \int \diff{x'} \rho(x') \nabla c_2(x, x'),
\end{align}
where $\Vert \cdot \Vert_\infty$ denotes the maximum norm.
Panels (a) and (b) of Fig.~\ref{fig:noether} depict results for $e_1$ and $e_2$ as a function of the mean density $\bar{\rho} = \int \diff{\vec{r}} \rho(\vec{r}) / V$ for all 150 density profiles of the test set, where $V$ denotes the volume of the system.
The small magnitudes of the observed error values indicate that the neural network satisfies the Noether identities \eqref{eq:Noether_correlation} and \eqref{eq:Noether_shift} to very high accuracy.
Outliers can be attributed mostly to the moderate numerical noise of the spatial gradients, see panel (III) in Fig.~\ref{fig:noether}, and are no hinderance in practical applications of the neural functional.

This confirmation demonstrates that our method transcends the neural network from a mere interpolation device of the simulation training data to a credible standalone theoretical object.
The fact that one is able to carry out consistent and performant functional calculus indeed renders $c_1^\star(x; [\rho])$ a neural-network-based density functional.
Besides functional differentiation, we show next that functional line integration acts as the inverse operation and provides access to the corresponding free energy.
Appendix \ref{appendix:c2_symmetry} gives further insight into the symmetry properties of $c_2^\star(x, x'; [\rho])$, which serve as a prerequisite for the existence of a generating excess free energy functional $\Fexc^\star[\rho]$; we recall Eq.~\eqref{eq:c1_Fexc}.

\subsubsection{Equation of state and free energy}
\label{sec:free_energy}

Although the machine learning procedure operates on the level of the one-body direct correlation function, the excess free energy $\Fexc[\rho]$ is accessible by functional line integration \cite{Brader2015}:
\begin{equation}
  \label{eq:Fexc_from_c1}
  \beta \Fexc[\rho] = - \int_0^1 \diff{\alpha} \int \diff{\vec{r}} \rho(\vec{r}) c_1(\vec{r}; [\rho_\alpha]).
\end{equation}
Here, $\rho_\alpha(\vec{r}) = \alpha \rho(\vec{r})$ is a sequence of density profiles that are linearly parametrized by $\alpha$ in the range $0 \leq \alpha \leq 1$.
The limits are $\rho_0(\vec{r}) = 0$ such that $\Fexc[0] = 0$, and $\rho_1(\vec{r}) = \rho(\vec{r})$, which is the target density profile that appears as the functional argument on the left hand side of Eq.~\eqref{eq:Fexc_from_c1}.
Other parametrizations of $\rho_\alpha(\vec{r})$ are conceivable but change the concrete form of Eq.~\eqref{eq:Fexc_from_c1}.
On the basis of $c_1^\star(x; [\rho])$, we implement Eq.~\eqref{eq:Fexc_from_c1} via $\beta \Fexc^\star[\rho] = - A \int_0^1 \diff{\alpha} \int \diff{x} \rho(x) c_1^\star(x; [\rho_\alpha])$ and evaluate the integrals numerically; as before $A$ denotes the lateral system area.

We first return to bulk systems and illustrate in the following three different routes towards obtaining the bulk equation of state from the neural network.
For this, we introduce the excess free energy density as $\psi_b(\rho_b) = \Fexc[\rho_b] / V$, where $V$ is the system volume.
From the neural functional, the excess free energy density $\psi_b^\star(\rho_b)$ can be acquired via $\Fexc^\star[\rho_b]$ from functional line integration along a path of bulk densities according to Eq.~\eqref{eq:Fexc_from_c1}.
Alternatively and equivalently, one can simply evaluate the neural direct correlation functional at bulk density $\rho_b$ and due to translational symmetry at arbitrary location (e.g.\ $x = 0$) such that $c_1^{b \star} = c_1^\star(0; [\rho_b])$.
Simplifying Eq.~\eqref{eq:c1_Fexc} in bulk reveals that
\begin{equation}
  \label{eq:psi_from_c1_0}
  \psi_b^{\star\prime}(\rho_b) = - k_B T c_1^{b \star},
\end{equation}
where the prime denotes the derivative with respect to the bulk density argument.
The excess free energy density $\psi_b^{\star}(\rho_b)$ follows from ordinary numerical integration across bulk densities up to the target value $\rho_b$.
The numerical accuracy to which both routes coincide serves as a further valuable consistency test.

\begin{figure}[tb]
  \centering
  \includegraphics{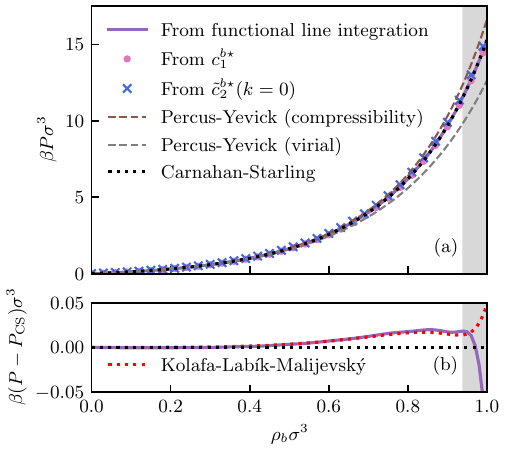}
  \caption{
    We show (a) the equation of state $P(\rho_b)$ obtained via different methods and (b) deviations to the Carnahan-Starling result $P_\mathrm{CS}(\rho_b)$ (dotted black line).
    The neural equation of state $P^\star(\rho_b)$ is calculated via Eq.~\eqref{eq:eos_psi} in which the excess free energy density follows from functional line integration according to Eq.~\eqref{eq:Fexc_from_c1} (solid purple line), from evaluation of the bulk value $c_1^{b\star}$ (pink dots), see Eq.~\eqref{eq:psi_from_c1_0}, and via the low-wavelength limit of $\tilde{c}_2^{b \star}(k)$ (blue crosses), see Eq.~\eqref{eq:eos_chi}.
    For comparison, the Percus-Yevick equations of state according to the virial (dashed gray line) and compressibility (dashed brown line) route are shown.
    Bulk densities $\rho_b$ beyond the stable fluid phase are shaded in gray.
    All three routes coincide very well up to and within the metastable region, with functional line integration leading to the most accurate results.
    In panel (b), we additionally depict a simulation-based equation of state (dotted red line) due to Kolafa, Labík and Malijevský \cite{Kolafa2004}, which our neural functional is able to reproduce very accurately in the stable fluid region, hence exceeding in precision the Carnahan-Starling equation of state.
  }
  \label{fig:free_energy_bulk}
\end{figure}

Additionally, one obtains the bulk pressure $P(\rho_b)$ from the excess free energy density via
\begin{equation}
  \label{eq:eos_psi}
  P(\rho_b) = \left( \psi_b^\prime(\rho_b) + k_B T \right) \rho_b - \psi_b(\rho_b).
\end{equation}
The pressure is equally accessible from a further route which incorporates previous results for the bulk pair structure via their low-wavelength limits according to \cite{Hansen2013}
\begin{equation}
  \label{eq:eos_chi}
  \beta \left. \frac{\partial P}{\partial \rho_b} \right\vert_T = \frac{\beta}{\rho_b \chi_T} = \frac{1}{S(0)} = \frac{1}{1 + \rho_b \tilde{h}(0)} = 1 - \rho_b \tilde{c}_2^b(0),
\end{equation}
where one can identify the isothermal compressibility $\chi_T = \rho_b^{-1} (\partial \rho_b / \partial P)_T$.
From Eq.~\eqref{eq:eos_chi}, $P(\rho_b)$ is obtained by evaluation of either of the bulk correlation functions (see Sec.~\ref{sec:twobody_bulk}) in Fourier space at $k = 0$ for different bulk densities and by subsequent numerical integration towards the target value of $\rho_b$.

We compare the results in Fig.~\ref{fig:free_energy_bulk}, where the equation of state $P^\star(\rho_b)$ of the neural network was acquired from functional line integration across bulk systems, cf.\ Eq.~\eqref{eq:Fexc_from_c1}, from evaluation of one-body bulk correlation values $c_1^{b \star}$, cf.\ Eq.~\eqref{eq:psi_from_c1_0}, and from the low-wavelength limit of two-body bulk correlations, cf.\ Eq.~\eqref{eq:eos_chi}.
One finds that the results of all three routes are consistent with each other and that they match very well the Carnahan-Starling equation of state \cite{Carnahan1969}, thus outperforming the Percus-Yevick theory as already observed for the bulk pair structure in Sec.~\ref{sec:twobody_bulk}.
A slight deviation can be noticed when evaluating $P^\star(\rho_b)$ via Eq.~\eqref{eq:eos_chi}, which constitutes the most indirect route detouring to two-body correlations.
This may reflect the small discrepancy of the neural functional to simulation results (cf.\ Fig.~\ref{fig:twobody_bulk_comparison}) and the sensitivity of the low-wavelength limit of the static structure factor to remaining finite size effects \cite{Hoefling2020}.
Notably, functional line integration is the most reliable method and the corresponding results even surpass the Carnahan-Starling equation of state in accuracy.
Fig.~\ref{fig:free_energy_bulk}(b) shows the reproduction of a highly accurate simulation-based equation of state due to Kolafa, Labík and Malijevský \cite{Kolafa2004}.
We recall again that neither bulk information nor data for free energies or pressures was given explicitly in the training of the neural network.
Instead, the beyond-Carnahan-Starling precision is achieved solely by extracting direct one-body correlations from simulation data of randomized inhomogeneous systems in planar geometry.
In Appendix \ref{appendix:dimensional_crossover}, we additionally demonstrate that the neural functional is fit for the application of dimensional crossover \cite{Rosenfeld1997} in order to obtain the bulk equation of state for the two-dimensional hard disk fluid within a reasonable range of packing fractions.

For a concise comparison of free energies in \emph{inhomogeneous} situations, additional reference data has to be acquired from simulations.
In our grand canonical setting, thermodynamic integration \cite{Frenkel2023} with respect to the chemical potential can be used to measure the grand potential according to
\begin{equation}
  \label{eq:thermodynamic_integration}
  \Omega[\rho] = - \int_{-\infty}^\mu \diff{\mu'} \langle N \rangle.
\end{equation}
Here, the integration starts from an empty system with $\Omega[0] = 0$ and traverses the chemical potential up to the target value $\mu$.
One needs to measure the mean number of particles $\langle N \rangle$ in a sufficient number of simulations with intermediate chemical potentials $-\infty < \mu' \leq \mu$ to evaluate Eq.~\eqref{eq:thermodynamic_integration} numerically.
The excess free energy then follows directly from
\begin{equation}
  \label{eq:Fexc_from_Omega}
    \Fexc[\rho] = \Omega[\rho] - F_\mathrm{id}[\rho] - \int \diff{\vec{r}} \rho(\vec{r}) (\Vext(\vec{r}) - \mu),
\end{equation}
where $F_\mathrm{id}[\rho] = k_B T \int \diff{\vec{r}} \rho(\vec{r}) (\ln \rho(\vec{r}) - 1)$ is the ideal gas free energy.
Thermodynamic integration according to Eq.~\eqref{eq:thermodynamic_integration} has been performed for 22 systems of the test set to yield reference values $\Fexc^\mathrm{sim}$ for the excess free energy via Eq.~\eqref{eq:Fexc_from_Omega}.
The systems were selected to cover a broad range of excess free energy values, and FMT results for $\Fexc$ were used as a further theoretical estimate for this selection.

\begin{figure}[tb]
  \centering
  \includegraphics{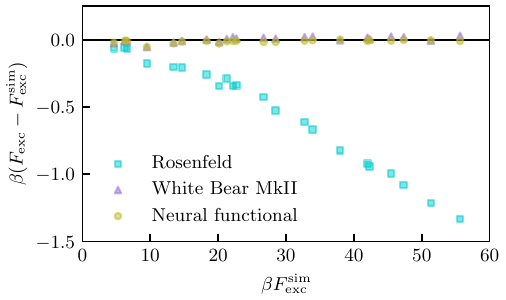}
  \caption{
    We compare free energies of inhomogeneous test systems as obtained via Rosenfeld (turquoise squares) and White Bear (purple triangles) FMT as well as with functional line integration of the neural correlation functional $c_1^\star(x; [\rho])$ (yellow circles).
    The discrepancy $\Fexc - \Fexc^\mathrm{sim}$ of the respective method to the simulation result $\Fexc^\mathrm{sim}$ is shown.
    Rosenfeld FMT systematically underestimates $\Fexc$ whereas White Bear MkII FMT as well as our neural functional yield almost exact results.
    The neural network performs slightly better for large excess free energies as occur primarily in dense systems.
  }
  \label{fig:free_energy}
\end{figure}

In Tab.~\ref{tab:free_energy} and Fig.~\ref{fig:free_energy}, we show errors of $\Fexc$ to the quasi-exact simulation values when calculating the excess free energy via Rosenfeld and White Bear MkII FMT as well as from functional line integration according to Eq.~\eqref{eq:Fexc_from_c1} of the neural functional.
For both FMT methods, a DFT minimization (cf.\ Sec.~\ref{sec:beyond_analytic}) is performed to yield a self-consistent density profile $\rho(x)$, which serves as input to the respective analytic FMT expression for $\Fexc[\rho]$.
Hence we compare consistently equilibrium states (according to the respective theory) corresponding to the same form of the external potential.

\begin{table}[tb]
  \caption{
    The absolute and relative mean average error of the excess free energy $\Fexc$ as obtained via the Rosenfeld and White Bear MkII FMT functionals is compared to the result from functional line integration of the neural correlation functional.
    The reference values $\Fexc^\mathrm{sim}$ were obtained via thermodynamic integration according to Eqs.~\eqref{eq:thermodynamic_integration} and \eqref{eq:Fexc_from_Omega} for a subset of the test systems.
    The results of the neural functional surpass the Rosenfeld FMT significantly and even yield a slight improvement over the highly accurate and refined White Bear theory.
    The angular brackets denote an average over the 22 test simulations.
  }
  \label{tab:free_energy}
  \centering
  \begin{tabular}{c|c|c}
    & $\beta \langle |\Fexc - \Fexc^\mathrm{sim}| \rangle$ & $\langle |\Fexc - \Fexc^\mathrm{sim}| / \Fexc^\mathrm{sim} \rangle$ \\
    Rosenfeld & 0.540 & 1.75\% \\
    White Bear MkII & 0.0159 & 0.104\% \\
    Neural functional & 0.0127 & 0.097\%
  \end{tabular}
\end{table}

The comparison reveals that the neural functional significantly outperforms Rosenfeld FMT and still yields slightly more accurate values for the excess free energy than the very reliable White Bear theory.
Regarding the above described bulk results for the free energy, this behavior is both consistent and expected, as the Rosenfeld and White Bear MkII functionals can be associated with the Percus-Yevick compressibility and Carnahan-Starling bulk equations of state respectively.
Still, the test in inhomogeneous systems is a more rigorous one than in bulk, as the full nonlocal functional representation is invoked when providing $c_1^\star(x; [\rho])$ with an inhomogeneous density profile as input.
Given that the functional line integration of $c_1^\star(x; [\rho])$ via Eq.~\eqref{eq:Fexc_from_c1} is practically immediate, one can deem $\Fexc^\star[\rho]$ itself a corresponding neural functional for the excess free energy that enables a full description of the thermodynamics of inhomogeneous fluids to high accuracy.
As we present below, this quantitative precision is preserved when applying the neural functional in a predictive manner in the self-consistent calculation of density profiles.

\subsection{Predicting inhomogeneous fluids via neural DFT}

\subsubsection{Going beyond analytic approximations}
\label{sec:beyond_analytic}

In the previous section, the trained model has been put to test by deriving related quantities such as $c_2^\star(x, x'; [\rho])$ from autodifferentiation and $\Fexc^\star[\rho]$ from functional line integration in order to assess its performance against analytic and numerical reference results.
We now turn to the application of the neural functional $c_1^\star(x; [\rho])$ in the context of the self-consistent determination of density profiles according to the DFT Euler-Lagrange equation.
This is achieved by rearranging Eq.~\eqref{eq:c1_EL} to the standard form \cite{Evans1979,Hansen2013}
\begin{equation}
  \label{eq:DFT_EL}
  \rho(\vec{r}) = \exp\left(-\beta (\Vext(\vec{r}) - \mu) + c_1(\vec{r}; [\rho])\right).
\end{equation}
A fixed-point (Picard) iteration with mixing parameter $\alpha$ can be used to determine the density profile from Eq.~\eqref{eq:DFT_EL} according to
\begin{equation}
  \label{eq:Picard_iteration}
  \begin{split}
    \rho(\vec{r}) &\leftarrow (1 - \alpha) \rho(\vec{r})\\
                  &\quad + \alpha \exp\left(-\beta (\Vext(\vec{r}) - \mu) + c_1(\vec{r}; [\rho])\right).
  \end{split}
\end{equation}
The degree of convergence is determined from the remaining difference of right and left hand side of Eq.~\eqref{eq:DFT_EL}.
With the trained neural functional at hand, one can evaluate the one-body direct correlation function in Eq.~\eqref{eq:Picard_iteration} via the surrogate $c_1^\star(x; [\rho])$ in each iteration step.
In the following, the use of $c_1^\star(x; [\rho])$ in this context will be referred to as neural DFT.

We note two minor technical points concerning the use of the neural functional in the Picard iteration.
It was observed that a conservative choice of $\alpha$ is necessary during the first few iterations to ensure numerical stability.
After this burn-in, the mixing parameter can be set to usual values (e.g.\ $\alpha = 0.05$).
Furthermore, the convergence criterion has to be relaxed as compared to typical choices in analytic DFT methods due to the remaining intrinsic uncertainty of $c_1^\star(x; [\rho])$.
The mean average error after training, cf.\ Sec.~\ref{sec:training}, provides an estimate for the expected relative uncertainty of the density profile according to Eq.~\eqref{eq:DFT_EL}.
Depending on the specific problem, the error might not decrease any further than that during the iteration of Eq.~\eqref{eq:Picard_iteration}.
Neither of these points caused any practical hinderance in applications.

The treatment of Eq.~\eqref{eq:DFT_EL} in neural DFT is conceptually not different than in standard DFT methods.
However, the model $c_1^\star(x; [\rho])$ relieves the theory from being restricted by the available approximations for the one-body direct correlation function as generated from analytic expressions of the excess free energy functional $\Fexc[\rho]$ via Eq.~\eqref{eq:c1_Fexc}.
We emphasize that, unlike in previous work \cite{Lin2019,Yatsyshin2022}, no analytic ansatz had to be provided and that our method is generic for the determination of a suitable functional from a given model Hamiltonian, thus indeed constituting a ``machine learning black box'' \cite{Lin2019} regarding the training procedure.
However, in contrast to a closed black box, the inner workings of the resulting neural correlation functional can be inspected very thoroughly via the neural functional calculus laid out above.
Also note that, while the model works at the level of the one-body direct correlation function, the free energy is readily available from functional line integration, cf.\ Sec.~\ref{sec:free_energy}.
Lastly, we point out that $c_1^\star(x; [\rho])$ captures the entirety of the intrinsic correlations and that further improvements are conceivable by only learning differences to an analytic reference functional.
To demonstrate the capabilities of our method, we refrain from this route and show that the trained neural functional alone already exceeds the accuracy of FMT.

\subsubsection{Comparison to FMT}
\label{sec:beyond_fmt}

\begin{figure}[!tb]
  \centering
  \includegraphics{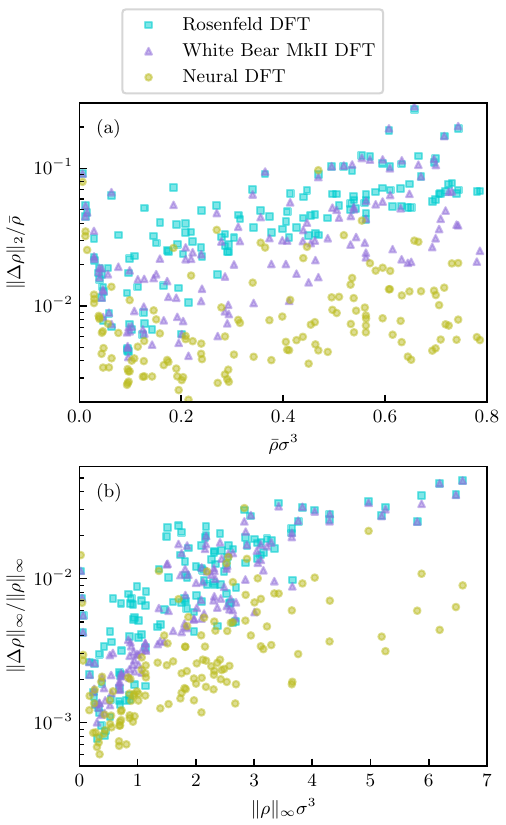}
  \caption{
    Measures of discrepancy of self-consistent density profiles to simulation results across the test set are presented.
    We show (a) the normalized $L_2$-norm $\Vert \Delta \rho \Vert_2 / \bar{\rho}$ as a function of the mean density $\bar{\rho}$ for judgment of the average error over the inhomogeneous system, and (b) the relative maximum norm $\Vert \Delta \rho \Vert_\infty / \Vert \rho \Vert_\infty$ as a function of the largest local density $\Vert \rho \Vert_\infty$ to reveal the magnitude of local errors, e.g.\ at density peaks.
    The self-consistent density profiles are obtained from Rosenfeld (turquoise squares) and White Bear MkII (purple triangles) FMT \cite{Rosenfeld1989,HansenGoos2006} as well as from employing our neural functional $c_1^\star(x; [\rho])$ in the DFT Euler-Langrange equation (yellow circles).
    Regarding both global and local error, the neural network outperforms the analytic FMT functionals and reduces the respective errors up to an order of magnitude, especially in large density regimes.
  }
  \label{fig:stats_scatter}
\end{figure}

In the following, we benchmark the self-consistent inhomogeneous density profiles obtained via neural DFT against FMT results.
For this comparison, the Rosenfeld \cite{Rosenfeld1989} and White Bear MkII \cite{HansenGoos2006} FMT functionals are considered and the simulated density profiles are taken as quasi-exact reference data.
The FMT functionals are the most profound analytic description of the hard sphere fluid with the White Bear MkII theory being the state-of-the-art treatment of short-ranged intermolecular repulsion in classical DFT.
Nevertheless, measurable and systematic deficiencies still remain, e.g.\ in highly correlated systems \cite{Davidchack2016}.
We point the reader to Ref.~\cite{Roth2010} for a thorough account of FMT and to Ref.\ \cite{Sammueller2023} for a very recent quantitative assessment.
Note that the tensorial weights of Tarazona \cite{Tarazona2000} to describe hard sphere freezing are not included in our investigation.

The comparison is set up as follows.
For each hard sphere system of the test set (see Sec.~\ref{sec:training}), we determine the density profile $\rho(x)$ from the Rosenfeld and White Bear MkII FMT functionals as well as from $c_1^\star(x; [\rho])$ via the Picard iteration \eqref{eq:Picard_iteration} of the Euler-Lagrange Eq.~\eqref{eq:DFT_EL}.
For this, only the known form of the external potential $\Vext(x)$ and the value $\mu$ of the chemical potential are prescribed.
As reference density profiles are available from GCMC simulations, we can evaluate the error $\Delta \rho(x)$ of each of the DFT results relative to the simulation data for $\rho(x)$.
From here, different scalar metrics for the quantitative agreement of self-consistent DFT profiles and simulation results are considered.

In Fig.~\ref{fig:stats_scatter}, both global and local error measures for the deviation of FMT as well as neural DFT to simulation data are depicted.
For the assessment of the global error, we show the $L_2$-norm $\Vert \Delta \rho \Vert_2$ of the discrepancy to the reference profile, which is normalized by the mean density $\bar{\rho}$ of each system respectively.
As the test data covers very dilute to very dense systems, this relative global error measure is plotted as a function of $\bar{\rho}$ to discern the behavior with respect to varying global average density.
Similarly, we define an estimate for the relative local error by evaluating the maximum norm $\Vert \Delta \rho \Vert_\infty$ of the density deviation divided by the maximum value $\Vert \rho \Vert_\infty$ of the GCMC density profile.
This quantity is resolved against the maximum $\Vert \rho \Vert_\infty$ of the respective inhomogeneous density, thus enabling the detection of local discrepancies, e.g.\ in the vicinity of maxima and discontinuities of the density profile.

One recognizes that neural DFT yields substantially better results than the FMT functionals with regard to both error measures.
Compared to the Rosenfeld results, both the global and the local error is decreased by approximately an order of magnitude.
Surprisingly, even the White-Bear MKII functional is not able to match the accuracy of the neural DFT, which is noticeable especially for large values of $\bar{\rho}$ and of $\Vert \rho \Vert_\infty$.

\subsubsection{Simulation beyond the box}
\label{sec:beyond_the_box}

\begin{figure*}[tb]
  \centering
  \includegraphics{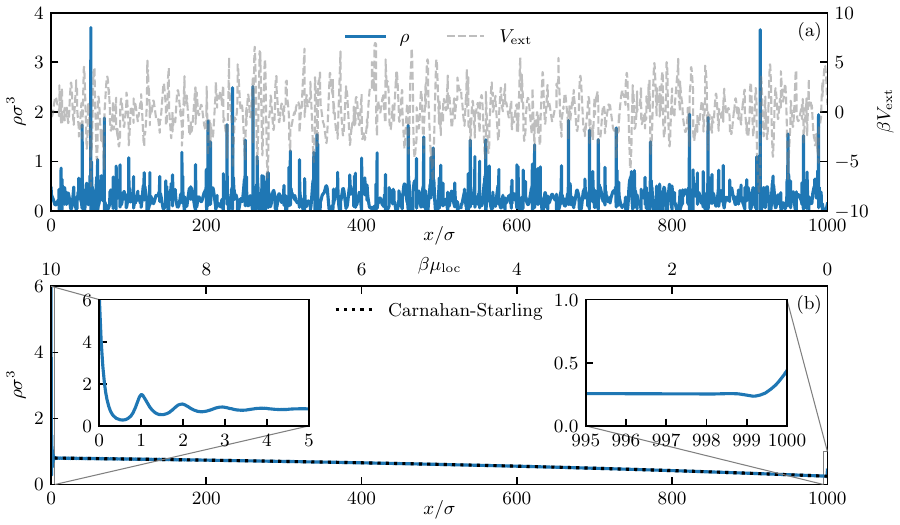}
  \caption{
    Neural DFT is used to obtain the density profile $\rho(x)$ (blue lines) of the hard sphere fluid (a) in a highly correlated system with randomized external potential $V_\mathrm{ext}(x)$ (gray dashed line) and (b) in a sedimentation column of height $1000 \sigma$ that is bounded by hard walls at the bottom and at the top.
    Near-simulation microscopic accuracy is retained at low computational cost by the application of neural DFT in the highly correlated large-scale system.
    For the case of sedimentation, strongly oscillating behavior at the lower wall as well as mild adsorption at the top can be resolved.
    As the spatial variation of the local chemical potential $\mu_\mathrm{loc}(x)$ is negligible, the density profile reproduces the equation of state within the sedimentation column, which is verified by a comparison to the Carnahan-Starling equation of state (dotted black line).
  }
  \label{fig:multiscale}
\end{figure*}

A particular advantage of the local nature of the neural functional $c_1^\star(x; [\rho])$ is its applicability to systems of virtually arbitrary size.
As explained in Sec.~\ref{sec:neural_network}, it is sufficient to provide the density profile within a rather narrow window as input to the neural network to infer the value of the one-body direct correlation function at the center of the density section.
The model $c_1^\star(x; [\rho])$ can therefore be used directly in the Euler-Lagrange Eq.~\eqref{eq:DFT_EL} for the prediction of planar systems of arbitrary length.
Due to the low computational demands of solving this equation self-consistently, this method is suitable even in multiscale problems where macroscopic length scales compete with and are influenced by microscopic correlations and packing features.
Although one could argue that analytic DFT methods already account for such tasks, importantly the neural functional $c_1^\star(x; [\rho])$ acts as a drop-in replica of the (almost) simulation-like description of the intrinsic correlations.
Therefore, neural DFT facilitates to fuse simulation data with common DFT methods, thus providing a means to ``simulate beyond the box''.

Simulation beyond the box is demonstrated in Fig.~\ref{fig:multiscale}, where the system size has been increased to $1000 \sigma$ while the numerical grid size remains unchanged at $0.01 \sigma$.
Our setup implies that for colloids of, say, size $\sigma = \SI{1}{\micro\meter}$, we have spatial resolution of $\SI{10}{\nano\meter}$ across the entirety of a system of macroscopic size $\SI{1}{\milli\meter}$.
We consider both a highly correlated fluid in a rapidly varying external potential as well as the diffusive sedimentation behavior \cite{Heras2013} in a weak gravitational potential.
The former case is realized by generating a sequence of randomized external potentials via Eq.~\eqref{eq:Vext} which are spatially connected; the chemical potential is set to zero.
Neural DFT yields a highly inhomogeneous density profile in this system and resolves the microscopic variations accurately at low computational cost.
In the sedimentation column, a local chemical potential $\mu_\mathrm{loc}(x) = \mu - \Vext(x) = (10 - 0.01 x / \sigma) k_B T$ is imposed which decreases linearly with respect to the height $x$, and the system is bounded from the bottom ($x = 0$) and the top ($x = 1000 \sigma$) by hard walls.
The spatial variation of $\mu_\mathrm{loc}(x)$ is chosen small enough to enable thermal diffusion across the whole sedimentation column and to yield locally an almost bulk-like behavior except near the upper and lower hard walls.
The method reproduces both the highly correlated nature of $\rho(x)$ in the vicinity of the walls as well as its intermediate behavior within the sedimentation column, which follows closely the bulk equation of state (see Sec.~\ref{sec:free_energy}), as one would expect within a local density approximation \cite{Hansen2013}.
In both cases, the computational cost for the determination of $\rho(x)$ with neural DFT is negligible as compared to analogous many-body simulations, which are hardly feasible on such length scales.

\section{Discussion}
\label{sec:conclusion}

In this work, we have outlined and validated a machine learning procedure for representing the local functional map from the density profile to the one-body direct correlation function via a neural network.
The resulting neural functional was shown to be applicable as a powerful surrogate in the description of inhomogeneous equilibrium fluids.
This was demonstrated for the hard sphere fluid, where we have used GCMC simulations in randomized inhomogeneous planar environments for the generation of training, validation and test data.
Density and one-body direct correlation profiles followed respectively from direct sampling and from evaluation of Eq.~\eqref{eq:c1_EL}.

DFT elevates the role of the one-body direct correlation function $c_1(x)$ to that of an intrinsic functional $c_1(x; [\rho])$ depending on the density profile $\rho(x)$ but being independent of the external potential.
We exploited this fact in the construction of our neural network, which takes as input a local section of the discretized density profile around a fixed location $x$ and outputs the value of the one-body direct correlation functional $c_1(x; [\rho])$ at that specific location.
Establishing a pointwise inference of $c_1(x; [\rho])$ instead of trying to represent the global functional mapping of the entire one-body profiles comes with various advantages, such as independence of the box size, the correct description of the short-range behavior of $c_1(x; [\rho])$, and a very significant improvement of training statistics.

The nonlinear and nonlocal functional relationship was realized by fully-connected hidden layers with smooth activation functions and a standard supervised training routine was used.
The achieved mean average error over the test set was of the same order of magnitude as the noise floor of the simulations, thus being indicative of full representational power of the neural correlation functional within the considered simulation data.
Whether the quality of the model can be improved further by performing more extensive sampling to reduce the statistical noise of the simulation profiles remains to be investigated in the future.
Additionally, active and reinforcement machine learning techniques could be useful for interleaving the training and simulation process, thereby guiding the generation of reference data in order to explore the space of inhomogeneous systems more efficiently and exhaustively.

The neural functional was put to test by verifying numerous physical relations in bulk and in inhomogeneous systems.
In particular, it was shown that the two-body direct correlation functional $c_2(x, x'; [\rho])$ as well as higher-order correlations are accessible from the model via automatic differentiation.
In bulk, the pair structure as described by the neural network significantly outperforms the Percus-Yevick theory and is even able to compete with simulation results \cite{Groot1987}, although no bulk data was used during training.
In inhomogeneous situations, the conformance of the neural functional to the thermal Noether sum rules \eqref{eq:Noether_correlation} and \eqref{eq:Noether_shift} as well as to spatial symmetry requirements holds to high accuracy.
The excess free energy $\Fexc[\rho]$ is readily and efficiently available via functional line integration of the model according to Eq.~\eqref{eq:Fexc_from_c1} and the results agree with those obtained from simulations.
The bulk equation of state can be acquired consistently from various routes with the results attaining simulation quality \cite{Kolafa2004} and in particular exceeding the very reliable Carnahan-Starling equation of state \cite{Carnahan1969} in accuracy.
Dimensional crossover is feasible for the calculation of the bulk equation of state for the two-dimensional hard disk system.

Arguably the most important consequence of the neural functional framework is the applicability of $c_1^\star(x; [\rho])$ in the self-consistent calculation of density profiles by solving the Euler-Lagrange Eq.~\eqref{eq:DFT_EL} of classical DFT.
As the one-body direct correlation function is faithfully represented by the neural network, one is exempted from having to find analytic approximations for $c_1(x; [\rho])$ or for its generating functional $\Fexc[\rho]$.
Although FMT provides such approximations for the hard sphere fluid with high precision, we could demonstrate that our neural functional outperforms both the Rosenfeld \cite{Rosenfeld1989} as well as the White Bear MkII \cite{HansenGoos2006} functional.
For this, Eq.~\eqref{eq:DFT_EL} was solved self-consistently for all 150 randomized local chemical potentials of the test set to obtain $\rho(x)$, where $c_1(x; [\rho])$ was given either analytically by FMT or evaluated via $c_1^\star(x; [\rho])$.
The comparison of the results to the simulated density profiles reveals that neural DFT yields global and local errors that are up to an order of magnitude lower than those of FMT.

Furthermore, due to the flexibility that comes with the local functional mapping, the neural network could be used as a means to ``simulate beyond the box''.
That is, while the training was based solely on simulation data from systems of manageable size, the resulting model $c_1^\star(x; [\rho])$ is directly applicable for predictions on much larger length scales.
We demonstrated this by imposing a spatial sequence of randomized external potentials on a length of $1000 \sigma$.
While the explicit numerical simulation of such a system is comparatively cumbersome, neural DFT offers a way to achieve close to simulation-like accuracy at low computational effort.
Furthermore, we have considered a sedimentation column with a height of $1000 \sigma$ that is bounded by hard walls.
Neural DFT is capable to both resolve microscopically the adsorption at the walls as well as to efficiently capture the long-range density decay with increasing height.
The presented fusion of machine learning and DFT can therefore be another useful technique to make headway in the multiscale description of soft matter \cite{Site2019,Baptista2021,Schmid2022}.

Even though we saw no need for a more sophisticated training procedure in our investigations, it could be useful to consider physics-informed machine learning \cite{Karniadakis2021} as a technique for enforcing exact physical relations of the underlying problem directly during training.
Sum rules in bulk or in inhomogeneous systems, e.g.\ the thermal Noether identities \eqref{eq:Noether_correlation} and \eqref{eq:Noether_shift}, might be suitable candidates for this task.
Analogous to the evaluation of derivatives in physics-informed neural networks, we have shown the necessary quantities to be accessible by automatic differentiation of the neural functional.

When considering nonequilibrium systems, power functional theory (PFT) \cite{Schmidt2013,Schmidt2022} establishes an exact functional many-body framework which is analogous to that of DFT in equilibrium.
A central ramification of PFT is the existence of a functional map from the time-dependent one-body density $\rho(\vec{r}, t)$ and current $\vec{J}(\vec{r}, t)$ to the internal force profile $\vec{f}_\mathrm{int}(\vec{r}, t; [\rho, \vec{J}])$, which is in general nonlocal in space and causal in time $t$.
Recent work by de las Heras \textit{et~al.}~\cite{Heras2023} demonstrated that machine learning this kinematic internal force functional yields highly promising results and overcomes the analytic and conceptual limitations of dynamical density functional theory.
In this regard, our method can be put into a more general context as it may be viewed as a mere special case for equilibrium systems where $\vec{J}(\vec{r}, t) = 0$.
The topical problem of accurately describing nonequilibrium many-body physics is certainly a natural contender for the application and extension of our neural functional framework, with many practical questions arising, e.g.\ concerning the generation of training data or the choice of neural network architecture.

While much insight could be gained by considering the hard sphere fluid, the application of our machine learning procedure is arguably even more useful for particle models that lack satisfactory analytic density functional approximations.
Although mean-field descriptions account surprisingly well for soft and attractive contributions \cite{Archer2017,Tschopp2020}, e.g.\ in the Lennard-Jones fluid, analytic efforts to go beyond this approximation are sparse \cite{Schmidt1999,Schmidt2000,Finster2022}.
We demonstrate the generality of our method in Appendix \ref{appendix:LJ}, where we show that the machine learning routine applies directly to the (truncated) Lennard-Jones interaction potential in an isothermal supercritical setting.
In the future, providing the temperature as a further input quantity to a modified neural network is a valuable goal in order to tackle the full physics of such thermal systems.
As a proper treatment of the arising phase transitions and interfacial phenomena is already subtle in simulation, the machine learning perspective might provide further insights.
We expect the general method to hold up even for complex particle models, e.g.\ containing many-body interactions \cite{Molinero2008}, provided that sufficiently accurate training data of sufficient quantity can be generated.

For the treatment of anisotropic particles, the neural network must be extended to accomodate for the additional orientational degrees of freedom.
Recent advances in molecular DFT could be helpful in guiding appropriate augmentations of our method \cite{Jeanmairet2013,Ding2017}.
Related to the increased dimensionality due to anisotropy, the extension of the machine learning procedure from planar symmetry to more general geometries is worth contemplating.
Especially for fully inhomogeneous three-dimensional problems, the amount of required training data seems restrictive at first.
However, we have shown in this work that results obtained in planar geometry already capture the essence of internal interactions.
Therefore, it may be feasible to base the machine learning predominantly on data in reduced geometrical settings and to incorporate remaining nontrivial effects due to the more general geometry by supplementing only a few selected higher-dimensional simulations.
In particular, we highlight in this context the promising development of equivariant neural networks \cite{Cohen2016,Weiler2018,Finzi2020,Satorras2021}, which serve as a means of casting underlying symmetries of a problem directly into the neural network architecture.
Recent applications in the physical domain show that this method facilitates robust training and generalization on the basis of much reduced data sets as compared to common machine learning approaches which do not intrinsically enforce symmetry \cite{Batzner2022,Batzner2023,Musaelian2023}.
In our case, exploiting inherent symmetries of the direct correlation functional via the use of equivariant neural networks is certainly valuable when further orientational or spatial degrees of freedom are to be considered.

Lastly, we point out useful cross-fertilization of machine learning ideas regarding topical applications in quantum DFT \cite{Pederson2022}.
In particular, the analogous functional mapping to the classical one-body direct correlation functional $c_1(\vec{r}; [\rho])$ is given quantum mechanically by the exchange-correlation potential $v_\mathrm{xc}(\vec{r}; [n])$ which depends functionally on the one-body electron density $n(\vec{r})$.
Due to the immediate analogy, obtaining the exchange-correlation energy functional $E_\mathrm{xc}[n]$ might be feasible with functional line integration similar to our treatment of $\Fexc[\rho]$ via Eq.~\eqref{eq:Fexc_from_c1}, which here becomes $E_\mathrm{xc}[n] = \int_0^1 \diff{\alpha} \int \diff{\vec{r}} n(\vec{r}) v_\mathrm{xc}(\vec{r}; [n_\alpha])$ with $n_\alpha(\vec{r}) = \alpha n(\vec{r})$.
Albeit lacking the neural functional calculus that we presented here, Zhou \textit{et al.}\ \cite{Zhou2019} have successfully demonstrated the machine learning of the functional mapping from the electron density to local values of the exchange-correlation potential $v_\mathrm{xc}(\vec{r})$.
Specifically, they trained a convolutional neural network on the basis of three-dimensional quantum chemical simulation data of small molecules and could obtain accurate predictions for larger molecules.
This success is akin to the multiscale applicability of our neural correlation functional $c_1^\star(\vec{r}; [\rho])$.
In general, however, most machine learning strategies in quantum DFT have considered different functional mappings \cite{Nagai2018,Schmidt2019,Nagai2020,Li2021,Li2022,Wang2023}.
In light of our results for classical systems, we deem the analogous machine learning of the local functional relationship of $v_\mathrm{xc}(\vec{r}; [n])$ the arguably most promising approach in the development of a neural quantum DFT with the goal of chemical accuracy and generic applicability.

\begin{acknowledgments}
  We thank T.\ Zimmermann, T.\ Eckert and N.\ C.\ X.\ Stuhlmüller for useful comments.
  This work is supported by the German Research Foundation (DFG) via Project No.\ 436306241.
\end{acknowledgments}

\section*{Data availability}

Code, simulation data and models that support the findings of this study have been deposited in Zenodo \cite{Zenodo}.

\bibliography{bibliography}

\appendix

\section{Higher-order correlations}
\label{appendix:c3}

\begin{figure*}[tb]
  \centering
  \includegraphics{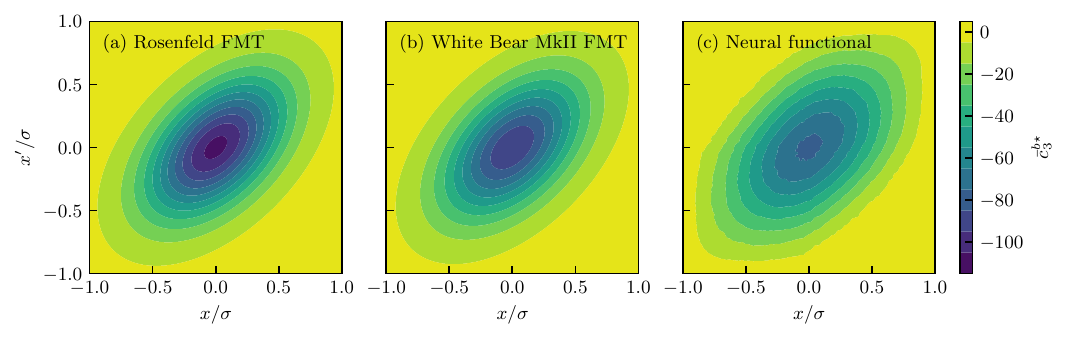}
  \caption{
    The three-body direct correlation function is shown in bulk at density $\rho_b = 0.7 \sigma^{-3}$.
    We depict (a) the Rosenfeld and (b) the White Bear MkII FMT results for the planar representation $\bar{c}_3^b(x, x')$, which were obtained analytically according to Eq.~\eqref{eq:c3_FMT} by a cumulant expansion in Fourier space and a subsequent backtransform.
    Within our neural functional framework (c), $\bar{c}_3^{b \star}(x, x')$ is acquired via nested automatic differentiation of $c_1^\star(x; [\rho])$.
  }
  \label{fig:c3}
\end{figure*}

Analogous to Sec.~\ref{sec:twobody_bulk}, we demonstrate that higher-order correlations can be obtained from the neural correlation functional by nested automatic differentiation.
This is due to the fact that the hierarchy of direct correlation functions $c_n(\vec{r}, \vec{r}', \dots, \vec{r}^{(n-1)}; [\rho])$, $n \geq 2$, is accessible from successive functional derivatives of the one-body direct correlation functional \cite{Hansen2013},
\begin{equation}
  \label{eq:cn}
  c_n(\vec{r}, \vec{r}', \dots, \vec{r}^{(n-1)}; [\rho]) = \frac{\delta^{n-1} c_1(\vec{r}; [\rho])}{\delta \rho(\vec{r}') \dots \delta \rho(\vec{r}^{(n-1)})}.
\end{equation}
As illustrated in the main text, translational symmetry can be applied in bulk fluids such that the resulting bulk correlation function $c_n^b(\vec{r}, \dots, \vec{r}^{(n-2)}) = c_n(0, \vec{r}, \dots, \vec{r}^{(n-2)}; [\rho_b])$ only incorporates $n - 2$ remaining position coordinates.

We specialize again to the planar geometry of our neural functional and show in Fig.~\ref{fig:c3} the three-body bulk correlation function $\bar{c}_3^{b \star}(x, x')$ for a bulk density of $\rho_b = 0.7 \sigma^{-3}$.
While the computation of $\bar{c}_2^{b \star}(x)$ is practically immediate via a single reverse mode autodifferentiation pass, going to the three-body correlation function comes at the price of having to evaluate the Hessian of $c_1^\star(x; [\rho])$, for which different strategies exist \cite{Dixon2008}.
In principle, one can proceed by nesting autodifferentiation layers to obtain further members of the hierarchy \eqref{eq:cn}, albeit being restricted by the practicability of the actual evaluation and the efficacy of the result.
Note that the computational effort at the three-body level is by no means restrictive and that growing numerical demands are expected when considering higher-order correlations.
The computation and analysis of $\bar{c}_3^{b}(x, x')$ might be especially useful for more complex fluid models, e.g.\ containing internal three-body interactions \cite{Molinero2008}.

We compare $\bar{c}_3^{b \star}(x, x')$ to analytic approximations based on FMT.
For both the Rosenfeld and the White Bear MkII functional, the three-body bulk direct correlation function is analytic in Fourier space.
We point the reader to Ref.~\cite{Rosenfeld1989} for an expression of the original Rosenfeld result in terms of vectorial weight functions and to Refs.~\cite{Kierlik1990,Phan1993} for an equivalent representation via scalar weights.
As the weight functions remain unchanged, the White Bear MkII result follows immediately from the modification of the excess free energy density as laid out in Ref.~\cite{HansenGoos2006}.

A cumulant expansion of the bulk result of the three-body direct correlation function in Fourier space can be transformed to real space analytically, which in planar geometry gives
\begin{equation}
  \label{eq:c3_FMT}
  \bar{c}_3^b(x, x') = - \frac{b R^4}{a} \exp\left(\frac{-x^2 + x x' - x'^2}{a R^2} \right),
\end{equation}
where the width parameter $a$ and the prefactor $b$ are determined by
\begin{align}
  a &= \frac{\nu}{\kappa} \frac{3}{5} \frac{53 - 25 \eta + 8 \eta^2}{30 + 2 \eta + 5 \eta^2 - \eta^3},\\
  b &= \kappa \frac{8 \pi}{3 \sqrt{3}} \frac{30 + 2 \eta + 5 \eta^2 - \eta^3}{(1 - \eta)^5},
\end{align}
with the packing fraction $\eta = \pi \rho_b / 6$.
The correction factors $\nu$ and $\kappa$ are set to unity in the Rosenfeld FMT and attain the forms
\begin{align}
  \nu &= \frac{53 - 35 \eta + \eta^2 + 5 \eta^3}{53 - 25 \eta + 8 \eta^2},\\
  \kappa &= \frac{30 - 6 \eta}{30 + 2 \eta + 5 \eta^2 - \eta^3},
\end{align}
in the White Bear MkII case.

The comparison reveals that the form of the neural three-body bulk correlation function $\bar{c}_3^{b \star}(x, x')$ is plausible and that it captures genuine features which go beyond both FMT descriptions.
The Rosenfeld FMT yields a large discrepancy in the core region $x, x' \approx 0$, which is significantly unterestimated as compared to the results from the neural functional and from the White Bear theory.
We recall that, as in Sec.~\ref{sec:beyond_fmt}, the tensorial weights of Tarazona \cite{Tarazona2000} have not been used in the FMT functionals and that their inclusion might be particularly relevant on the level of higher-order correlations.
In this vein, investigating members of the direct correlation hierarchy \eqref{eq:cn} with the neural correlation functional could be a valuable aid for testing and refining analytic FMT functionals.

\section{Spatial symmetry of the neural two-body direct correlation functional}
\label{appendix:c2_symmetry}

\begin{figure}[tb]
  \centering
  \includegraphics{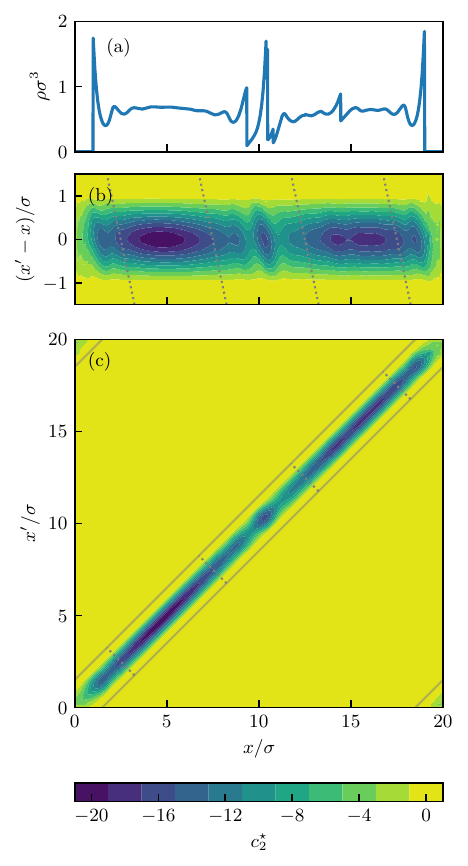}
  \caption{
    We show (a) the density profile $\rho(x)$ of an inhomogeneous system of the test set and (b) the corresponding neural two-body direct correlation function, which is obtained for each position $x$ with respect to $x' - x$.
    A linear transformation is applied to display $c_2^\star(x, x'; [\rho])$ as a function of $x$ and $x'$ in panel (c).
    This transformation is visualized by corresponding gray lines in panels (b) and (c) which indicate the extent of the detailed view (solid) and slices where $x + x' = \mathrm{const.}$ (dotted).
    The results exemplify that the neural network reproduces the symmetry property \eqref{eq:c2_symmetry} of the two-body direct correlation function very accurately.
  }
  \label{fig:c2_symmetry}
\end{figure}

A further consistency test of $c_2^\star(x, x'; [\rho])$ arises due to its expected symmetry with respect to an interchange of the planar position coordinates $x$ and $x'$.
Recall that the excess free energy functional $\Fexc[\rho]$ generates the two-body direct correlation function according to
\begin{equation}
  \label{eq:c2_from_Fexc}
  c_2(\vec{r}, \vec{r}'; [\rho]) = - \frac{\delta^2 \beta \Fexc[\rho]}{\delta \rho(\vec{r}) \delta \rho(\vec{r}')},
\end{equation}
see Eqs.~\eqref{eq:c1_Fexc} and \eqref{eq:c2} of the main text.
One can directly recognize from the symmetry of the second functional derivative in Eq.~\eqref{eq:c2_from_Fexc} that $c_2(\vec{r}, \vec{r}'; [\rho]) = c_2(\vec{r}', \vec{r}; [\rho])$ must hold.

On the basis of the neural direct correlation functional in planar geometry, assessing the validity of the identity
\begin{equation}
  \label{eq:c2_symmetry}
  c_2^\star(x, x'; [\rho]) = c_2^\star(x', x; [\rho])
\end{equation}
is a highly nontrivial test.
This is due to the fact that $c_2^\star(x, x'; [\rho])$ evaluated at certain positions $x$ and $x'$ follows from automatic differentiation of $c_1^\star(x; [\rho])$, where the input density window is centered around the location $x$, see Sec.~\ref{sec:twobody_bulk}.
On the other hand, when formally evaluating $c_2^\star(x', x; [\rho])$, where the arguments $x$ and $x'$ are now reversed, the density window is centered around $x'$, hence constituting a generally very different and a priori unrelated input profile.
One can expect Eq.~\eqref{eq:c2_symmetry} to be recovered only if the physical implications of Eq.~\eqref{eq:c2_from_Fexc} are captured correctly by the neural functional.
Note that Eq.~\eqref{eq:c2_symmetry} is a necessary condition for the existence of a unique neural excess free energy functional $\Fexc^\star[\rho]$, which can practically be obtained via functional line integration, see Sec.~\ref{sec:free_energy}.
We exemplify in Fig.~\ref{fig:c2_symmetry} that the neural two-body direct correlation functional $c_2^\star(x, x'; [\rho])$ obtained via autodifferentiation of $c_1^\star(x; [\rho])$ indeed satisfies the symmetry requirement \eqref{eq:c2_symmetry} to very high accuracy.

\section{Neural equation of state for hard disks via dimensional crossover}
\label{appendix:dimensional_crossover}

\begin{figure}[htb]
  \centering
  \includegraphics{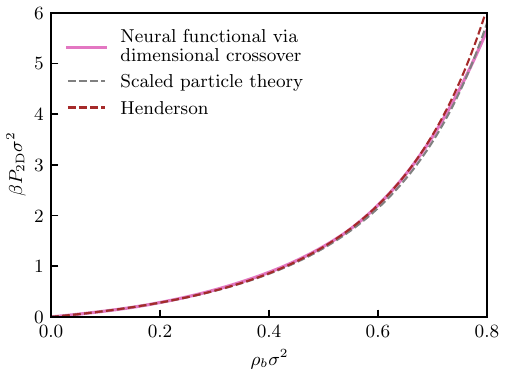}
  \caption{
    The equation of state $P_\mathrm{2D}(\rho_b)$ for two-dimensional hard disks is depicted, which is obtained from the neural functional via dimensional crossover.
    For comparison, we show analytic results according to scaled particle theory \cite{Reiss1959} and by Henderson \cite{Henderson1975a}.
    Although the training data for the three-dimensional hard sphere fluid did not cover narrow confinement within hard walls, $c_1^\star(0; [\rho])$ reproduces very reasonable behavior when applied to such quasi-two-dimensional situations and yields acceptable results for densities up to $\rho_b \approx 0.7 \sigma^{-2}$.
  }
  \label{fig:dimensional_crossover}
\end{figure}

Although the neural functional $c_1^\star(x; [\rho])$ was acquired explicitly for the three-dimensional hard sphere fluid, dimensional crossover techniques can be used to obtain bulk results for the two-dimensional hard disk system.
This is facilitated by investigating the behavior of the hard sphere fluid under narrow confinement, which constitutes a quasi-two-dimensional scenario.
With this method, one obtains the equation of state for the hard disk fluid from $c_1^\star(x; [\rho])$, as we demonstrate in the following.

We proceed similar to Sec.~\ref{sec:free_energy} and utilize Eq.~\eqref{eq:eos_psi} to express the pressure $P(\rho_b)$ via the excess free energy density $\psi_b(\rho_b)$, which we aim to compute for a range of bulk densities $\rho_b$.
Whereas $c_1^\star(x; [\rho])$ was evaluated for the three-dimensional bulk fluid at spatially constant density, cf.\ Eq.~\eqref{eq:psi_from_c1_0}, here a suitable density profile $\rho_\mathrm{2D}(x)$ is constructed as input to the neural direct correlation functional in order to emulate narrow planar confinement.
For this, we choose
\begin{equation}
  \label{eq:rho_2D}
  \rho_\mathrm{2D}(x) = \frac{\rho_b}{x_w} \Theta\left(\left|x - \frac{x_w}{2}\right|\right)
\end{equation}
with the Heaviside function $\Theta(\cdot)$; note that Eq.~\eqref{eq:rho_2D} is a Dirac series and yields the Dirac distribution for $x_w \rightarrow 0$.
The neural direct correlation functional is then evaluated at the center of this assumed slit, and the values $c_1^\star(0; [\rho_\mathrm{2D}])$ are used analogous to Sec.~\ref{sec:free_energy} for the determination of $P_\mathrm{2D}^\star(\rho_b)$.
The equation of state for the associated two-dimensional hard disk system follows formally for $x_w \rightarrow 0$.
As this limit is not directly accessible in practice, we assess the obtained values for finite but small slit widths $0.3 \leq x_w / \sigma \leq 1$ and extrapolate to $x_w = 0$ via a quadratic fit.

The resulting equation of state $P_\mathrm{2D}^\star(\rho_b)$ for the two-dimensional hard disk fluid as obtained from this dimensional crossover on the basis of the neural network is shown in Fig.~\ref{fig:dimensional_crossover}.
We additionally display analytic equations of state from scaled particle theory \cite{Reiss1959} and by Henderson \cite{Henderson1975a} which serve as reference.
One recognizes that reasonable results can be achieved for low and medium densities, but that deviations to analytic results become noticeable for $\rho_b > 0.7 \sigma^{-2}$.
Nevertheless, it is both surprising and reassuring that the neural functional is capable of predicting correlations in narrow confinement, as no such situations were explicitly included in the training data.
Recall that hard walls were imposed only at the borders of the simulation box of length $L = 20 \sigma$ and that the inhomogeneous external potential within the simulation domain consisted solely of Fourier modes and of piecewise linear functions, cf.\ Eq.~\eqref{eq:Vext} in the main text.
Presumably, improvements over the results presented in Fig.~\ref{fig:dimensional_crossover} could be obtained especially for large densities by including situations of very narrow confinement explicitly in the training data.
From our outset, the successful achievement of a viable two-dimensional equation of state serves as a demonstration that $c_1^\star(x; [\rho])$ indeed captures the intricate functional relationship of the underlying physical problem instead of acting as a mere interpolation tool with respect to the encountered training data.

\section{Neural DFT for the Lennard-Jones fluid}
\label{appendix:LJ}

\begin{figure}[tb]
  \centering
  \includegraphics{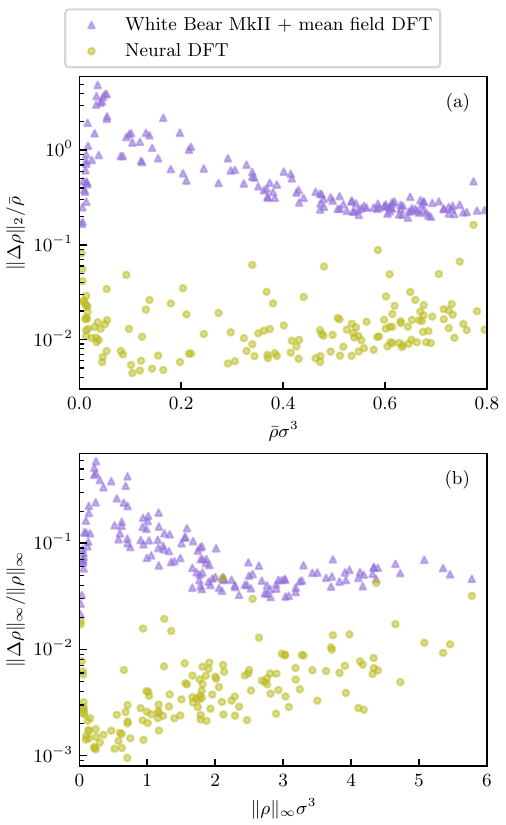}
  \caption{
    Neural DFT (yellow circles) is compared to the standard mean field DFT (purple triangles) for the truncated Lennard-Jones fluid.
    As in Fig.~6, (a) the normalized $L_2$-norm $\Vert \Delta \rho \Vert_2 / \bar{\rho}$ as a function of the mean density $\bar{\rho}$, and (b) the relative maximum norm $\Vert \Delta \rho \Vert_\infty / \Vert \rho \Vert_\infty$ as a function of the largest local density $\Vert \rho \Vert_\infty$ are considered.
    While considerable deviations to the reference profiles are observed for the hard sphere plus mean field treatment, neural DFT achieves almost simulation-like accuracy with global and local errors being decreased by up to two orders of magnitude.
  }
  \label{fig:stats_LJ_scatter}
\end{figure}

We illustrate the generalizability of our machine learning framework to other particle types by considering the truncated Lennard-Jones fluid with pairwise interparticle potential
\begin{equation}
  \label{eq:LJ}
  \phi(r) = \begin{cases}
              4 \epsilon \left[ \left( \frac{\sigma}{r} \right)^{12} - \left( \frac{\sigma}{r} \right)^{6} \right], & r \leq r_c, \\
              0, & r > r_c,
            \end{cases}
\end{equation}
where $r$ is the interparticle distance, $\epsilon$ is the dispersion energy and the cutoff radius is set to $r_c = 2.5\sigma$.
Analogous to Sec.~\ref{sec:simulation} of the main text, reference data is generated via GCMC simulations of 800 systems with randomized external conditions of which 500 are used for training and 150 respectively for validation and testing.
We focus on the isothermal behavior of the supercritical fluid and hence set $k_B T = 1.5 \epsilon$.
The chemical potential varies uniformly in a range of $-8 \leq \beta \mu \leq 4$ and the external potential is generated as described in the main text, cf.\ Eq.~\eqref{eq:Vext}.

To accomodate the longer-ranged interactions compared to the hard sphere fluid, the size of the density window to be input into the neural network is increased to $x_c = 4 \sigma$ whilst keeping the design of the hidden layers unchanged (see Sec.~\ref{sec:neural_network}).
The training results in a mean average error of $0.0035$ and larger values of $x_c$ led to no further improvement in the training statistics.
The slight increase of the mean average error as compared to the hard sphere case (see Sec.~\ref{sec:training}) can be attributed to noisier simulation data, which results from the decreased efficiency of GCMC method when simulating soft interactions with larger cutoff radius instead of hard spherical particles with an interaction range of $\sigma$.

After successfully training the neural functional for the Lennard-Jones fluid, we employ $c_1^\star(x; [\rho])$ in neural DFT to determine self-consistent density profiles for all 150 test systems.
The Picard iteration proceeds without problems and analogously to Sec.~\ref{sec:beyond_analytic}.
The results are compared with the standard mean field DFT treatment of the Lennard-Jones fluid.
Here, the repulsive part of Eq.~\eqref{eq:LJ} is approximated by a hard core interaction, for which we utilize the White Bear MkII FMT functional.
An additive mean field contribution $F_\mathrm{MF}[\rho] = \int \diff{\vec{r}} \int \diff{\vec{r}'} \rho(\vec{r}) \rho(\vec{r}') \phi_\mathrm{att}(|\vec{r} - \vec{r}'|) / 2$ to the excess free energy functional incorporates the attractive part $\phi_\mathrm{att}(r)$ of the Lennard-Jones potential.
The function $\phi_\mathrm{att}(r)$ is equal to Eq.~\eqref{eq:LJ} for $r \geq r_\mathrm{min} = 2^{1/6} \sigma$ and it is set to $-\epsilon$ for $r < r_\mathrm{min}$.

Local and global deviations of both neural DFT and the analytic mean field DFT to the simulation reference data are presented in Fig.~\ref{fig:stats_LJ_scatter}.
The neglection of correlations in the mean field treatment leads to considerable errors across the whole test set.
Contrarily, the neural DFT achieves close-to-simulation results and outperforms the analytic DFT by up to two orders of magnitude in the considered error measures.

This successful test demonstrates the transferability of our machine learning framework across particle models and indicates its utility especially for Hamiltonians which lack satisfactory analytic DFT treatments.
Although the considered interparticle potential \eqref{eq:LJ} is still short-ranged, we see much potential to extend our method to long-ranged interactions as occur e.g.\ in charged systems.
The resulting algebraic decay of direct correlations could be tackled in various ways:
i) It might be sufficient in some cases (e.g.\ for screened interactions) to simply extend the cutoff range $x_c$ of the density input.
ii) In order to achieve a better scaling of the number of input nodes with growing $x_c$, one could change the corresponding discretization of $\rho(x)$ to employ variably spaced sampling points instead of a fixed discretization interval.
This would still enable to finely resolve the vicinity of the considered location $x$ while also incorporating information about long-range density correlations.
iii) An alternative approach emerges by treating the long-range behavior of $c_1(\vec{r}; [\rho])$ analytically, similar to the treatment of the Hartree term in quantum DFT, see e.g.\ Ref.~\cite{Zhou2019}.
Hence, the neural functional could be trained as is on the remaining short-ranged part of $c_1(\vec{r}; [\rho])$ to recover full quasi-exact information about intrinsic correlations.

\end{document}